\documentclass[aoas,MSNbibl,nameyear,seceqn,dvips]{arximspdf}
\usepackage{dcolumn}
\usepackage{graphicx}

\doi{10.1214/12-AOAS556} 
\volume{6}
\issue{4}
\pubyear{2012}
\firstpage{1730}
\lastpage{1774}

\makeatletter
\newcommand{\eqref}[1]{(\ref{#1})}
\newcolumntype{d}[1]{D{.}{.}{#1}}
\makeatother

\begin{document}
\begin{frontmatter}

\title{Optimal obstacle placement with disambiguations}
\runtitle{Optimal obstacle placement}

\begin{aug}
\author[A]{\fnms{Vural}~\snm{Aksakalli}\corref{}\ead[label=e1]{aksakalli@sehir.edu.tr}}
\and
\author[B]{\fnms{Elvan}~\snm{Ceyhan}\ead[label=e2]{elceyhan@ku.edu.tr}}
\runauthor{V. Aksakall{\i} and E. Ceyhan}
\affiliation{Istanbul \c{S}ehir University and Ko\c{c} University}
\address[A]{Department of Industrial Engineering\\
Istanbul \c{S}ehir University\\
Istanbul 34662\\
Turkey\\
\printead{e1}}

\address[B]{Department of Mathematics\\
Ko\c{c} University\\
Istanbul 34450\\
Turkey\\
\printead{e2}}
\end{aug}

\received{\smonth{1} \syear{2012}}

%
\begin{abstract}
We introduce the optimal obstacle placement with disambiguations
problem wherein the goal is to place true obstacles in an
environment cluttered with false obstacles so as to maximize the
total traversal length of a navigating agent (NAVA). Prior to the
traversal, the NAVA is given location information and probabilistic
estimates of each disk-shaped hindrance (hereinafter referred to as
disk) being a true obstacle. The NAVA can disambiguate a disk's
status only when situated on its boundary. There exists an obstacle
placing agent (OPA) that locates obstacles prior to the NAVA's traversal.
The goal of the OPA is to place true obstacles in between the clutter
in such
a way that the NAVA's traversal length is maximized in a game-theoretic
sense. We assume the OPA knows the clutter spatial distribution type,
but not the exact locations of clutter disks. We analyze the traversal
length using repeated measures analysis of variance for various
obstacle number, obstacle placing scheme and clutter spatial
distribution type combinations in order to identify the optimal
combination. Our results indicate that as the clutter becomes more
regular (clustered), the NAVA's traversal length gets longer
(shorter). On the other hand, the traversal length tends to follow a
concave-down trend as the number of obstacles increases. We also
provide a case study on a real-world maritime minefield data set.
\end{abstract}

%

\end{frontmatter}

\section{Introduction}
A challenging stochastic optimization problem that has practical applications
in robotics, computer vision and naval logistics is the
\textit{stochastic obstacle scene \textup{(}SOS\textup{)} problem}.
This problem was
first introduced by \citet{pap91}, and its graph-theoretic version
was called the \textit{Canadian traveler's problem}.
Both continuous and graph-theoretic versions of the problem have gained
considerable attention recently [see, e.g., \citet{nik08}, \citet
{xu09}, \citet{lik09},
\citet{eye09}, \citet{rdp3}].
In this article, we consider a slightly modified version of
the original SOS problem.
In this version, a point-sized navigating
agent (NAVA) needs to quickly traverse from a given starting point
to a target point through an arrangement of disk-shaped regions
(these regions shall be referred to as ``disks'' henceforth for brevity).
Some of these disks are true obstacles placed by another
agent, called the \textit{obstacle-placing agent} (\textit{OPA}), and the
rest is clutter, that is, false obstacles.
For instance, in case of a
naval logistics application, the true obstacles would be mines, and
the clutter could be rocks, metal pieces, debris, etc.
The OPA places the
true obstacles in between the clutter prior to the NAVA's traversal.
At the outset, the NAVA does not know the actual status of any disk.
However, the NAVA is given respective probabilities of each disk
being a true obstacle or a clutter.
Over the course of the
traversal, the NAVA has the option to disambiguate any (ambiguous)
disk, that is, learn with 100\% accuracy if it is a true obstacle.
This disambiguation can be performed when the NAVA is situated on a
disk's boundary.
The NAVA can pass through a disk only if a
disambiguation reveals that it is clutter, that is, not a true obstacle.
We assume that there is no limit on the number of
available disambiguations, and that disambiguations can be executed
only at a cost added to the overall length of the traversal.
We also
assume that the obstacle scene is static, that is, the disks do not
change location during the traversal, and the obstacle/clutter
status of a disk never changes.
The NAVA's challenge is to decide
what and where to disambiguate en route so as to minimize the total
length of the traversal.
This problem is called the SOS problem.
The OPA's challenge, on the other hand, is to place a given number of
true obstacles in between the clutter so as to maximize the
traversal length of the NAVA in a game-theoretic sense.
We call this
problem the \textit{obstacle placement with disambiguations
problem}, or the \textit{OPD problem} in short.

As discussed in \citet{bao07}, the SOS problem can be cast as a
Markov decision process, though with exponentially many states.
There are no efficiently computable optimal policies known for the
SOS problem, and many similar problems have been shown to be
intractable [\citet{pap91}, \citet{pro03}]. Nonetheless, several efficient
heuristics have been proposed for the problem; see, for example,
\citet{rdp2} and \citet{rdp3}.
In particular, the reset disambiguation
(RD) algorithm of \citet{rdp3} is an efficient heuristic for the SOS
problem in a continuous setting. This algorithm is provably optimal
for a restricted class of SOS problems, and it has been shown to
perform well for general instances of the problem.

Algorithms in the literature for the SOS problem and its variants---both
in continuous and discrete settings---have assumed that the
spatial distribution of possible-obstacles is given. In general,
performance of these algorithms has been evaluated under complete
spatial randomness assumption for both true obstacles and clutter.
In a broader scheme, there has been some research on detecting
(true) obstacles via the obstacle field's spectral image
properties [\citet{pri97}, \citet{ols03}] as well as the field's
spatial point
pattern characteristics [\citet{cre00}, \citet{cre01}, \citet{mus95},
\citet{wal02}]. These
studies, on the other hand, assume that the obstacle field's spatial
point distribution is given. To our knowledge, the important problem
of placing the obstacles to maximize a NAVA's total
traversal length, that is, the OPD problem, has not been studied
before.

The goal of this article is to introduce the OPD problem
and study the problem in one particular setting to
stimulate further research on this subject and lay ground for more
comprehensive prospective studies. In particular, this study is
limited to an investigation of relative efficiency of a variety
of obstacle placement schemes against different background clutter types
sampled from various spatial point distributions. Our goal is to
gain insight into which obstacle placement scheme works better for
which clutter type, and explore the effect of the number of
obstacles on the NAVA's traversal length. In particular, we would like
to address the following two critical research questions:
\begin{itemize}
\item
Given a clutter type,
what is the optimal number of obstacles and the obstacle pattern to use
so as to maximize the NAVA's total traversal length?
\item Obstacles are likely to be costly, and the OPA might not have
enough numbers of obstacles.
In this case, what is the optimal way to place a given number of
obstacles for a given clutter type?
\end{itemize}

The primary analysis tool we use is repeated measures analysis of
variance (ANOVA). Our specific setup leads to a three-way
repeated measures ANOVA problem where the treatment factors are as
follows:\vspace*{6pt}

\textit{Clutter type}: We consider 6 different point processes for
sampling clutter disk centers:
homogeneous and inhomogeneous Poisson processes, Mat\'{e}rn and
Thomas clustered point processes, and hardcore and Strauss regular
point processes.

\textit{Number of obstacles}: We consider 5 different numbers of
obstacles (20, 30, 40, 50, and 60, resp.).

\textit{Obstacle layout scheme}: We experiment with a total of 19
different obstacle placement patterns.
These patterns are sampled from a homogeneous Poisson process within
four different window forms: the clutter sampling window itself,
linear, V-, and W-shaped polygonal windows.\vspace*{6pt}

The response variable here is the total traversal length
of the NAVA including the cost of disambiguations. Without loss of
generality, we assume a fixed radius for both obstacle and clutter
disks. For computational efficiency, we work with an 8-adjacency
integer lattice discretization of the continuous setting as
in \citet{rdp3}. As for the NAVA's navigation algorithm, we use a simple
adaptation of the RD algorithm for the lattice discretization, which
we call the \textit{adapted RD} (\textit{ARD}) algorithm.

The rest of this manuscript is organized as follows:
The SOS problem is formally defined in Section~\ref{secprobdefn}
and
the ARD algorithm is outlined in Section~\ref{secard}.
The clutter spatial point distributions (i.e., clutter types) are
described in Section~\ref{secclutter},
and the obstacle placement patterns are introduced in
Section~\ref{secobst}. The experimental setup and the statistical analysis
of our Monte Carlo simulations are provided in
Section~\ref{secanalysis}.
In Section~\ref{seccobra} we illustrate our approach on a real-world
U.S. Navy minefield data set.
Summary, conclusions, and directions for
prospective research are presented in Section~\ref{secdisc}.

\section{The SOS and OPD problems}
\label{secprobdefn} 

The continuous SOS problem is formally defined as follows: Consider
a bounded obstacle field $\Omega\subset\mathbb{R}^2$. There exists
a clutter spatial point process $\mathcal{C}$ that generates points
$X_C \subset\Omega$ at which clutter disks are centered.
Next, an
obstacle-placing agent (OPA) samples disk centers $X_O \subset
\Omega$ from an obstacle spatial point process $\mathcal{O}$ and
places obstacle disks centered at $X_O$. A navigating agent (NAVA)
wishing to traverse from a given starting point $s \in\Omega$ and a
target point $t \in\Omega$ is equipped with a sensor
that assigns random marks $\rho_C\dvtx X_C \rightarrow(0,1]$ and
$\rho_O\dvtx X_O \rightarrow(0,1]$ prior to the NAVA's traversal.
When observing a realization of these processes, the NAVA only sees
$X:=X_C \cup X_O$. We assume
that, for all $x \in X$, $\rho(x)$ is the probability that $x \in
X_O$, that is, $x$ is a true obstacle. For every disk center $x$, the
possibly-obstacle disk $D_x$ is an open region with a fixed radius
$r>0$. The NAVA seeks to traverse a continuous $s,t$ curve in
$(\bigcup_{x \in X_O}D_x)^c$ of shortest achievable arc length,
where $A^c$ stands for complement of $A$.
We further suppose that there is a dynamic learning capability;
specifically, for all $x \in X$, when the curve is on the boundary
$\partial D_x$, the agent has the option to \textit{disambiguate} $x$,
that is, to learn $x \in X_O$ or not, but at a cost $c>0$ added
to the length of the curve. The NAVA can pass through disks that have
been disambiguated and found to be clutter, but needs to avoid
ambiguous disks as well as disks that have been disambiguated and found
to be a true obstacle. How the NAVA should route the continuous $s,t$
traversal curve---and where and when the disambiguations should be
performed---to minimize the length of this curve is called the \textit
{continuous SOS problem}.

The problem of placing the obstacles so as to maximize the NAVA's
traversal length in the SOS problem is called the \textit{OPD
problem}.
In this study, we consider a particular variant of the OPD
problem where the OPA knows the clutter spatial point distribution
(called clutter type for brevity),
but not the exact locations of the clutter disks. The motivation for
this variant is that the clutter location information requires
specific knowledge of the NAVA's sensor technology, which is not
necessarily accessible to the OPA. Nonetheless, it is still likely
that the OPA has information on the \textit{spatial distribution} of
the clutter disks. For instance, rock or debris distribution along a
specific coastline might follow a certain spatial point distribution
that is known to the OPA. We leave it to future research to study a
second variant where the OPA knows the exact locations of the
clutter disks.

For computational efficiency, we consider a discrete approximation
of the continuous setting on a subgraph of the 8-adjacency integer
lattice as in \citet{rdp3}. Specifically, this discretization is the
graph $\mathsf G$ whose vertices are all of the pairs of integers
$i,j$ such that $1 \leq i \leq i_{\mathrm{max}}$ and $1 \leq j \leq
j_{\mathrm{max}}$, where $i_{\mathrm{max}}$ and $j_{\mathrm{max}}$
are given integers. There are edges between all pairs of the
following four types of vertices: (1) $(i,j)$ and $(i+1,j)$ with
unit length, (2) $(i,j)$ and $(i,j+1)$ with unit length, (3) $(i,j)$
and $(i+1,j+1)$ with length $\sqrt{2}$, and (4) $(i+1,j)$ and
$(i,j+1)$ with length $\sqrt{2}$. One vertex in $\mathsf G$ is
designated as the starting point~$s$, another vertex in $\mathsf G$
is designated as the target point $t$. The NAVA is to traverse
from $s$ to $t$ in $\mathsf G$, only using edges that do not
intersect any true obstacles or ambiguous disks. If an edge
intersects any ambiguous disk, then a disambiguation of the obstacle
may be performed from either of the edge's endpoints that is outside
of the disk. As before, the goal is to devise an algorithm that
minimizes the expected length of the traversal by effective
exploitation of the disambiguation capability. We call this
discretization the \textit{discretized SOS problem}, which, in
effect, is a special case of the \textit{Canadian Traveler's
Problem} in the literature with statistical dependency among the
edges. The reader is referred to \citet{rdp2} for a review of the
literature that includes the history and development of the problems
that fall under the SOS problem umbrella.

\section{Adaptation of the reset disambiguation algorithm for the
discretized SOS problem}
\label{secard}
The reset disambiguation (RD) algorithm introduced
in \citet{rdp3} for the continuous SOS problem is a high performing
heuristic that is provably optimal for a related problem and also
optimal for a restricted class of instances for the original SOS
problem. Otherwise, the algorithm is generally suboptimal, but it is
both effective and efficiently computable. This algorithm can be
adapted to the discretized SOS problem as follows: We first define
the edge weight function below for each edge in~$\mathsf G$:
%
\begin{equation}\label{eqnwe}
w(e) := \ell(e) + \frac{1}{2} \sum_{i=1}^{|X|} \mathbf{1}_{e \cap D_i \ne
\varnothing} \biggl(\frac{c}{1-\rho_i}\biggr),
\end{equation}
where $\ell(e)$ is the Euclidean length of the edge (which
is either 1 or $\sqrt{2}$), and $\mathbf{1}$ is the indicator function
(taking value 1 or 0 depending on whether its subscripted expression
is true or false). For instance, weight of an edge not intersecting
any disks is equal to its Euclidean length. On the other hand,
weight of an edge intersecting a single disk is equal to the sum of
the edge's Euclidean length and the cost of disambiguation divided
by the probability that the disk is not an obstacle. The \textit{adapted RD \textup{(}ARD\textup{)} algorithm}
would then have the NAVA do the
following:

\begin{longlist}[(1)]
\item[(1)] Find the shortest $s,t$ walk in $\mathsf G$ with respect to the
edge weights specified by~\eqref{eqnwe}
(using, e.g., Dijkstra's algorithm). Start from $s$ and traverse this
walk until its first ambiguous edge $e$
is encountered at vertex $v$, with edge $e$ intersecting disk~$D_i$.
Notice that the NAVA might revisit a vertex over the course of the
traversal, making the NAVA's final trajectory a walk (and not a path).
\item[(2)] At this point (since the NAVA cannot enter an ambiguous disk)
disambiguate the disk $D_i$.
\item[(3)] Either remove disk $D_i$'s center point $X_i$ from $X$ or set
$\rho_i:=1$ depending on whether $D_i$ was just discovered to be,
respectively, a clutter or an obstacle.
\item[(4)] Repeat this procedure using $v$ as the new $s$ until $t$ is reached.
\end{longlist}

\begin{figure}
\centering
\begin{tabular}{@{}cc@{}}

\includegraphics{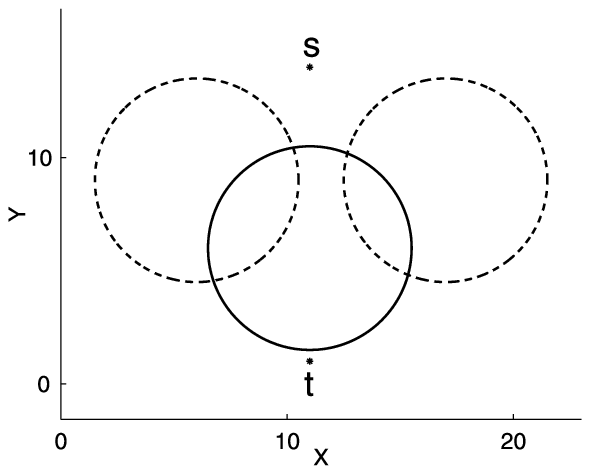}
 & \includegraphics{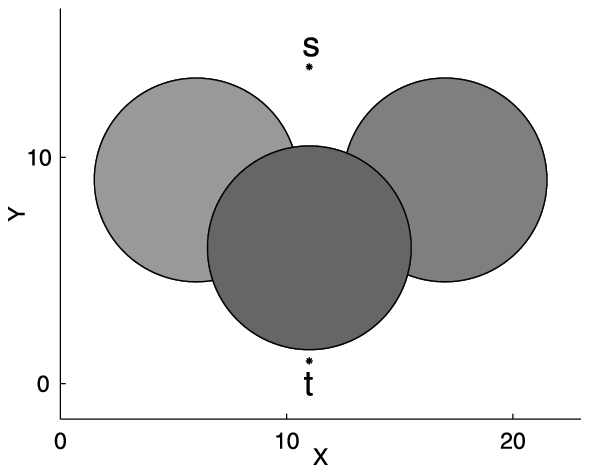}\\
\footnotesize{(a) The obstacle field} & \footnotesize{(b) The obstacle
field as seen by the NAVA}\vspace*{3pt}
\end{tabular}
\centering
\begin{tabular}{@{}c@{}}

\includegraphics{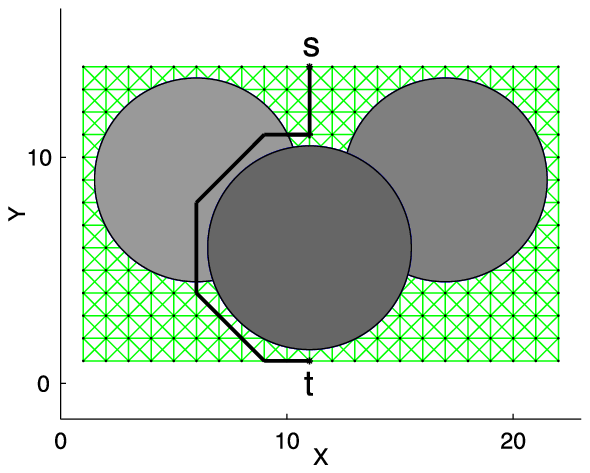}
\\
\footnotesize{(c) NAVA's traversal}
\end{tabular}
\caption{A simple setting with three disks and the NAVA's
traversal as dictated by the ARD algorithm.
\textup{(a)} shows the
actual status of the obstacle field. The two clutter disks on the
sides are denoted by dashed circles and the true obstacle in the
middle is denoted by a solid circle.
\textup{(b)} illustrates how the NAVA
sees the obstacle field prior to the navigation. Gray scale of disks
reflects marks of each disk, with darker colors indicating a higher
mark (for being a true obstacle).
\textup{(c)} shows the lattice discretization and the NAVA's actual
traversal.}\label{figexample}
\end{figure}

Figure~\ref{figexample} illustrates the SOS and OPD problem
settings with three disks. Two of these disks are clutter and the
third one is a true obstacle placed by the OPA. Clutter disks are
centered on the sides at $(6,9)$ and $(17,9)$, and the obstacle disk
is centered in the middle at $(11,6)$. The clutter disks will be
referred to as $D_1$ and~$D_2$, respectively, and the obstacle disk
will be referred to as $D_3$. It is important to reiterate that in
the variant we consider, the OPA knows the clutter distribution type,
but not the exact locations of clutter disks. In this specific
example, the OPA does not know where $D_1$ and $D_2$ are located, but
simply chooses to place $D_3$ at $(11,6)$. Each disk has a radius of
4.5 units and the cost of disambiguation is taken as 5 units. Actual
status of the obstacle field is shown in Figure~\ref{figexample}(a)
where clutter disks are shown as dashed circles and the true
obstacle is shown as a solid circle. The NAVA knows locations of all the
disks a priori, but does not know which disks are clutter and which
ones are truly obstacles. Instead, the NAVA is equipped with sensor
technology that assigns respective probabilities to each disk being
a true obstacle. These marks for $D_1, D_2$ and $D_3$ are taken as
0.4, 0.5 and~0.6, respectively. Shown in Figure~\ref{figexample}(b)
is how the NAVA sees the obstacle field where gray scale of disks
reflects the probability of each disk being a true obstacle as
measured via the NAVA's sensors prior to the NAVA's traversal, with darker
colors indicating higher probabilities of being a true obstacle.
Figure~\ref{figexample}(c) shows our lattice discretization and the
NAVA's actual traversal. Here, the lattice used is $22 \times14$,
with $s=(11,14)$ and $t=(11,1)$. The ARD algorithm first dictates
that $D_3$ is disambiguated at $(11,11)$. Since this is a true
obstacle, $\rho_1$ is set to 1 and the algorithm is queried again.
This time, the algorithm dictates that $D_1$ is disambiguated, again
at $(11,11)$. Since $D_1$ is clutter, the NAVA passes through it while
avoiding $D_3$ and reaches the target point. The NAVA's total traversal
length in this case is 29.49 including the cost of the two disambiguations.

In our computational experiments, the lattice used is
$(i_{\max} \times j_{\max})=(100 \times100)$, with $s=(50,100)$
and $t=(50,1)$.
Each disk has a radius of $r=4.5$ units, and the
disk centers are sampled on the pairs of real numbers in $[10,90]
\times[10,90]$---ensuring that there is always a permissible walk
from $s$ to $t$.
The cost of disambiguation is taken as $c=5$.
As in \citet{rdp1}, clutter marks are sampled from $\operatorname{Beta}(6,2)$ (with a
mean of 0.25) and obstacle marks are sampled from $\operatorname{Beta}(2,6)$ (with a
mean of
0.75). This particular setup has been specifically designed to possess
similar characteristics to an actual U.S. Navy minefield data set, called
the COBRA data, which was presented in \citet{wit95} and later used in
\citet{rdp2}, \citet{ye10}, and \citet{ye11}. In Section~\ref{seccobra}
we present an extensive case study on the COBRA data itself.\vadjust{\goodbreak}

\section{Clutter point distributions}
\label{secclutter} Formally, a \textit{spatial point process
}$X$ is a finite random subset of a bounded region $\Omega\subset
\mathbb{R}^2$.
A realization of this point process, on the other
hand, is called a \textit{spatial point pattern}. Classical
literature on the subject mainly identifies three spatial point pattern
categories based on the nature of inter-point interactions: (1)
independent patterns, (2) cluster patterns where points tend to be
close to one another, and (3) regular patterns where points tend to
avoid each other [\citet{bad10}]. In this study, we consider two
patterns from each one of these three categories in turn, and this
section describes those six spatial point processes used to generate
background clutter disk centers in the OPD problem. Classical
treatments of general spatial point patterns can be found
in \citet{cre93} and \citet{rip04}. The reader is referred
to \citet{mol07} for a brief overview of spatial point processes,
and to \citet{bad10} for an excellent coverage of the particular
point processes considered in this work.\looseness=-1

\subsection{Homogeneous and inhomogeneous Poisson processes}
\label{secH-and-IPP} In the context of spatial point processes,
intensity is the average density of points per unit area in the
region over which the point process is defined. In general, the null
model in a point pattern analysis is the \textit{homogeneous Poisson
point process} in the plane with constant intensity $\lambda$, which
is also called \textit{complete spatial randomness} (\textit{CSR}). CSR with
intensity $\lambda$ will be denoted by $\operatorname{CSR}(\lambda)$.
For any finite region $R$,
the CSR point process has four properties:
(1) the number of points in $R$ is a Poisson random variable,
(2) the number of points in any two disjoint regions $R$ and $R'$ are
independent random variables, (3) the expected number of points in
$R$ is $\lambda\cdot\operatorname{area}(R)$, and (4) the points in $R$ are
independently and uniformly distributed.

In the OPD problem, a possible scenario is that the density of the
clutter increases from the start point toward the target point
or vice versa. For instance, the density of rocks and/or debris
along a coast line might increase as one traverses toward the
coast. To simulate such a scenario, we consider the
\textit{inhomogeneous Poisson process}. This process is a
modification of CSR where the intensity is not constant, but varies
from location to location. Specifically, the intensity is a function
in two-dimensional Euclidean space. Let
$\operatorname{IP}(\lambda(h))$ denote the inhomogeneous Poisson
process with intensity $\lambda(h)$ where $h \in\mathbb{R}^2$.
Here, the intensity function $\lambda(h)$ specifies the values of
$\lambda$ on the plane. Properties of
$\operatorname{IP}(\lambda(h))$ are the same as those of
$\operatorname{CSR}(\lambda)$ with the last two properties modified
as follows: ($3'$) the expected number of points in $R$ is $\int_{R}
\lambda(h) \,dh$, and ($4'$) points in $R$ are independently and
identically distributed with probability density $f(h) = \lambda(h)
[\int_{R} \lambda(h) \,dh ]^{-1}$.

\subsection{\texorpdfstring{Mat\'{e}rn and Thomas clustered point processes}{Matern and Thomas clustered point
processes}}
In many real-world contexts, existence of a point at a specific
location increases the probability of other points being located in
its vicinity, giving rise to a \textit{clustered point
process}.\vadjust{\goodbreak}
Some examples include human settlements, plants, stars, galaxies
and molecules [\citet{dal02}]. In particular, it might be more
realistic to model clutter type in the OPD problem, such as
rock formation and debris dispersal along a coastline, as a
clustered point process rather than CSR.

A commonly-encountered cluster point process model in the literature
is the \textit{doubly-stochastic Poisson process}, which is also
known as the \textit{Cox process}. This process is a generalization
of the Poisson process where the intensity parameter is
randomized [\citet{dal02}]. In this work, we consider two special
cases of the Cox process: \textit{Mat\'{e}rn} and
\textit{Thomas point processes}. 

The Mat\'{e}rn point process, denoted $M(\lambda,\mu,r)$,
is constructed by first generating a
Poisson point process of ``parent'' points with intensity $\lambda$.
Each parent point is then replaced by a random cluster of points
where the number of points in each cluster is sampled from a Poisson
distribution with parameter $\mu$. These child points are placed
independently and uniformly inside a disk with a fixed radius, $r$,
centered at the parent point.

Similar to the Mat\'{e}rn point process, the Thomas process, denoted
$T(\lambda,\break\mu,\sigma)$, is
constructed by first generating a Poisson point process of ``parent''
points with intensity $\lambda$. A random cluster of points replaces
each parent point with the number of points per cluster being
sampled from a Poisson distribution with parameter $\mu$. In
contrast with the Mat\'{e}rn point process, positions of these child
points in the Thomas point process are isotropic Gaussian
displacements centered at the cluster parent location
with standard deviation~$\sigma$.

\subsection{Hardcore and Strauss regular point processes}
Another potential scenario in the OPD problem is where there is a
``regularity'' to the clutter disks.
That is, the clutter center
points tend to be a certain distance away from the other clutter
center points.
We consider two regular spatial point patterns with
pairwise interactions: the \textit{hardcore} and \textit{Strauss
point processes}.
The probability density function of the hardcore
process is that of the Poisson process with intensity $\lambda$
conditioned on the event that no two points generated by the
process are closer than $d$ units apart,
hence denoted at $\operatorname{HC}(\lambda,d)$.
The Strauss process, denoted $S(\lambda,d,\gamma)$, on
the other hand, generalizes the hardcore process by incorporating a
$\gamma\in[0,1]$ parameter that controls the interaction
between the points.
The process exhibits more regularity for smaller values of $\gamma$,
and less regularity for larger $\gamma$.
For $\gamma= 0$,
the Strauss process becomes a hardcore process, and for $\gamma= 1$,
it reduces to CSR [\mbox{\citet{bad10}}].

\subsection{The Clutter sampling procedure}
In our computational experiments, all spatial point processes---both
clutter and obstacle---are simulated via the \texttt{spatstat} package
in the R programming environment [\citet{bad05}].\vadjust{\goodbreak} This particular
package assumes that the point processes extend throughout the
two-dimensional Euclidean space, but they are observed inside a
sampling window $P$. In our case, the sampling window for the
clutter center points is taken as $P = [10,90] \times[10,90]$. In
sampling of the inhomogeneous Poisson process, points are generated
so that clutter density increases from the top of the obstacle field
toward the bottom where the target is located. Specifically,
the intensity function is taken as $\lambda(x,y) = 0.037 e^{(10-y)/40}$
on the sampling window $P$, which results in 100 points on the average.

In sampling of the Mat\'{e}rn and Thomas point processes, we work with
$M(10,10,10)$ and $T(10,10,5)$, respectively. As for the hardcore and
Strauss processes, we sample from $\operatorname{HC}(100,5)$ and $S(100,5,0.5)$. We
use the Metro\-polis--Hastings algorithm while sampling from the
hardcore and Strauss processes, which is essentially a Markov chain
whose states are spatial point patterns and its limiting
distribution is the desired point process. After running the
algorithm for a large number of times, which is 100,000 iterations
in our experiments, the state of the algorithm is considered to be a
realization of the desired point process [\citet{bad10}].

Parameter values of all the six clutter types are chosen
such that the number of points in any sampled point realization would be
roughly 100 on the average. For instance, CSR is sampled with number
of points being Poisson(100), and the $h(x,y)$ function we use for
the inhomogeneous Poisson process results in 100 points on
the average. However, the actual number of points in each clutter
realization that we use in our experiments is taken to be exactly 100.
This is
achieved by
rejection sampling, that is, by discarding sampled realizations for which
number of points is different than 100. The benefit of fixing the
number of clutter disks is that variation in traversal lengths
resulting from the different number of clutter disks is removed.
Thus, the only source
of variation in the background clutter in our computational
experiments is the spatial distribution of these 100 clutter disks.

\begin{figure}
\centering
\begin{tabular}{@{}cc@{}}

\includegraphics{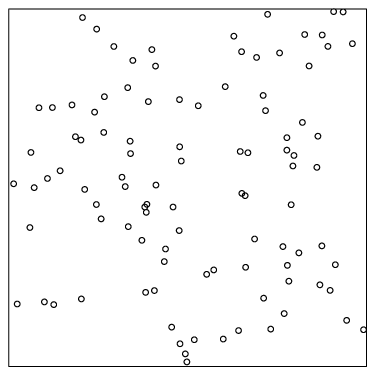}
 & \includegraphics{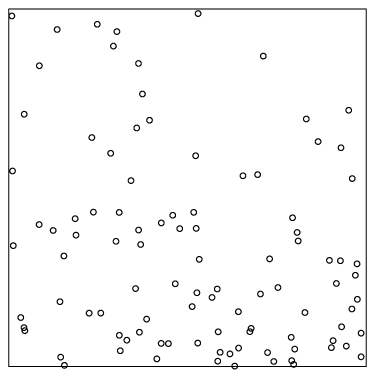}\\
\footnotesize{(a) CSR} & \footnotesize{(b) Inhomogeneous Poisson}\\[6pt]

\includegraphics{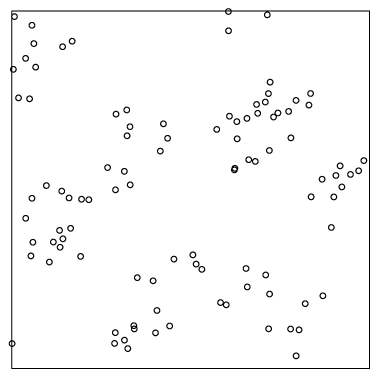}
 & \includegraphics{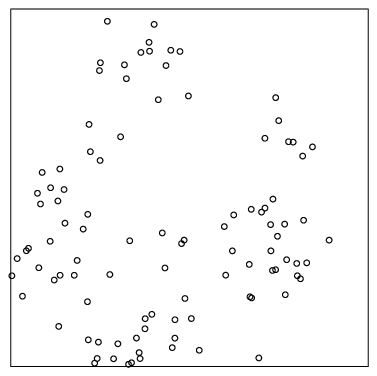}\\
\footnotesize{(c) Mat\'{e}rn} & \footnotesize{(d) Thomas}\\[6pt]

\includegraphics{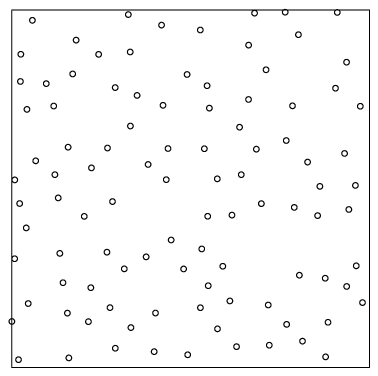}
 & \includegraphics{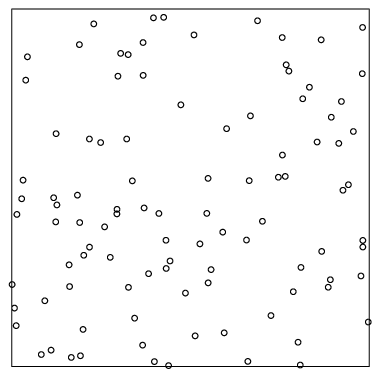}\\
\footnotesize{(e) Hardcore} & \footnotesize{(f) Strauss}\\
\end{tabular}
\caption{Sample realizations from the
six background clutter spatial point distributions.
The specific distribution parameters used are as follows: \textup{(a)}
$\operatorname{CSR}(100)$, \textup{(b)}
$\operatorname{IP}(0.037 e^{(10-y)/40})$, \textup{(c)}~$M(10,10,10)$, \textup{(d)}
$T(10,10,5)$, \textup{(e)}
$\operatorname{HC}(100,5)$ and \textup{(f)} $S(100,5,0.5)$. These parameters are chosen
such that the number of points in any sampled point realization would be
about 100 on the average. Rejection sampling was then utilized to
have exactly 100 points in all the clutter realizations.}
\label{figbg}
\end{figure}

It should be noted that using rejection sampling to achieve 100
points for each point pattern that we simulate actually changes the
distribution of the process which has produced the point pattern:
these points are now from a process conditional on the number of points
in the
region being 100.
In particular, the homogeneous Poisson process, that is,
CSR, conditioned on a specific number of points is in fact a uniform
distribution of that many points inside the sampling window.
However, the crucial observation here is that these conditional
processes share the same interaction behavior as the unconditional
ones, and this is sufficient for the purpose of our paper.
Figure~\ref{figbg} illustrates sample realizations from the clutter
point processes within our simulation environment.

%

\section{Obstacle placement schemes}
\label{secobst}
As mentioned earlier, the goal of the OPA is to
place a certain number of true obstacles in the obstacle field under
the assumption that the OPA knows the spatial point distribution of the
background clutter disks,\vadjust{\goodbreak} but not their exact locations. On the
other hand, the NAVA only has probabilistic information of each disk
being a true obstacle. The NAVA, however, can distinguish true
obstacles from the clutter only when situated at a disk's boundary.
In this study, we limit our focus to CSR for sampling true obstacle
disk centers within a total of 19
different sampling windows. One might also consider inhomogeneous
Poisson process for sampling the obstacle disk centers within these
sampling windows. In fact, increasing the obstacle intensity along
the $s-t$ line might perhaps increase the NAVA's traversal length in
general. However, we limit our focus to CSR for sampling the
obstacle disk centers for the following reasons: First, the OPA
might not know the exact starting and target points of the NAVA in
practice. Second, the area in which the OPA wishes to place obstacles
might not be a square region as in our experiments, but perhaps an
entire coastline. Thus, it makes more sense from an operational
point of view to sample the true obstacle disk centers with uniform
intensity inside their respective polygons.

We consider a total of 19 different sampling windows for the obstacle
disk centers.
The first sampling window is the polygon $P =
[10,90] \times[10,90]$, that is, the sampling window for the
background clutter. For the remaining windows, we consider 80-unit
long and 10-unit wide polygons as described below:\looseness=-1
\begin{itemize}
\item 8 different linear windows with their top left corner
$y$-coordinate being $90,80,\ldots,20$, and
\item 5 different V-shaped and W-shaped windows, respectively, with
their top left corner $y$-coordinate being $90,80,\ldots,50$.
The difference between the top and bottom $y$-coordinates of each one
of these 10 polygons is taken as 50 units.
\end{itemize}
The obstacle sampling window coinciding with the background clutter
window itself is code-named as P. Other sampling windows are code-named
by the polygon type (``L,'' ``V'' or ``W'', resp.) followed by
the top left corner coordinate of the polygon. These 4 polygon
shapes will be referred to as \textit{obstacle forms}.

For example, L70 is the polygon whose four corner points are
$(10,70)$, $(90,70)$, $(90,60)$ and $(10,60)$ clock-wise starting with the
top left corner. The polygon V70's six corner points are $(10,70)$,
$(50,40)$, $(90,70)$, $(90,60)$, $(50,30)$ and $(10,60)$, again clock-wise
starting with the top left corner. Similarly, polygon W70's ten
corner points are $(10,70)$, $(30,40)$, $(50,70)$, $(70,40)$, $(90,70)$,
$(90,60)$, $(70,30)$, $(50,60)$, $(30,30)$ and $(10,60)$.
The polygon W50,
for instance, is the same as W70 shifted down 20 units along the
$y$-axis. Thus, the 19 obstacle sampling windows we consider are P,
L90, L80$,\ldots,$L20, V90, V80$, \ldots,$V50, and W90, W80$, \ldots,$W50.
The reason we
consider the same polygon shape placed at different $y$-coordinates
is that we are not only interested in which polygon shape is more
efficient (in terms of increasing the NAVA's total traversal length),
but also which $y$-coordinate (i.e., distance to the target) is
more efficient for a given polygon shape. We do not consider placing
true obstacles along a straight horizontal line, as detection of
such obstacle patterns by the NAVA would be relatively
straightforward---see, for example, \citet{mus95} and \citet{wal02}
for detecting obstacles laid in such linear patterns.

\begin{figure}
\centering
\begin{tabular}{@{}ccc@{}}

\includegraphics{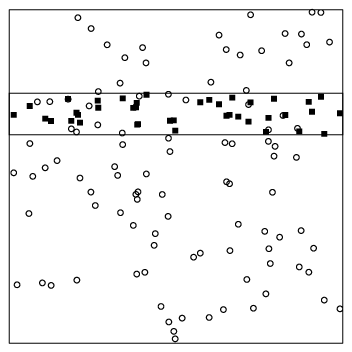}
 & \includegraphics{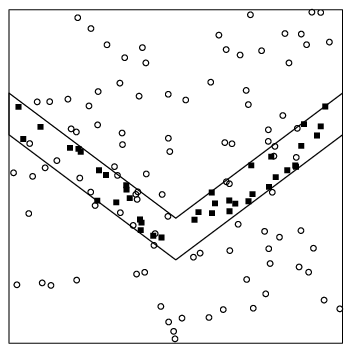}& \includegraphics{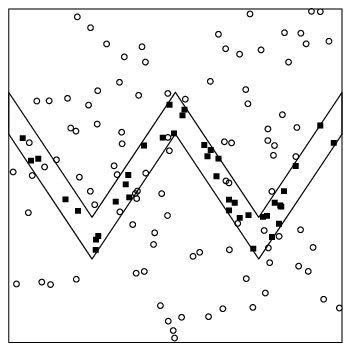}\\
\footnotesize{(a) CSR clutter${} + {}$L70:40} & \footnotesize{(b) CSR
clutter${} +
{}$V70:40}&\footnotesize{(c) CSR clutter${} + {}$W70:40}\\
\end{tabular}
\caption{Realizations of obstacle center points
inside polygons (indicated by solid boundaries)
L70, V70 and W70 with 40 obstacles superimposed on the CSR
clutter realization of Figure~\protect\ref{figexample}\textup{(a)}. Clutter disk centers are
denoted by
open circles ($\circ$) and obstacle disk centers are denoted by
solid squares~($\blacksquare$).}
\label{figobst}
\end{figure}

In order to assess the impact of the number of true obstacles in the
OPD problem, we consider five different number of obstacles for each\vadjust{\goodbreak}
one of the 19 obstacle sampling windows: 20, 30, 40, 50 and 60.
As mentioned earlier, CSR conditioned on a specific number of points is
in fact a uniform distribution with that many points. Therefore, the
obstacles we consider are essentially uniformly distributed inside
their respective sampling windows. Realizations of obstacle patterns are
denoted by the sampling window
followed by the number of obstacles. For instance, P:40 and L70:40
refer to the obstacle patterns sampled within the $P$ and L70
windows, respectively, with 40 obstacle center points inside their respective
windows. Figure~\ref{figobst} illustrates sample obstacle center
point realizations within the L70, V70 and W70 polygons, respectively,
with 40 obstacle center points within each polygon against the
CSR clutter shown in Figure~\ref{figbg}(a).

%

\begin{figure}
\centering
\begin{tabular}{@{}c@{}}

\includegraphics{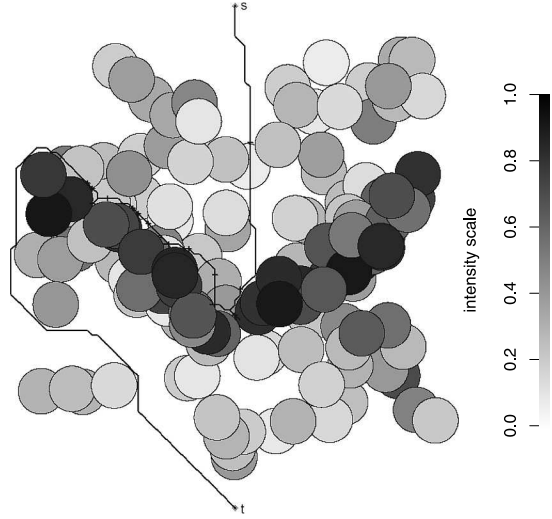}
\\
\footnotesize{(a) Navigation in CSR clutter${} + {}$V70:40}\\[6pt]

\includegraphics{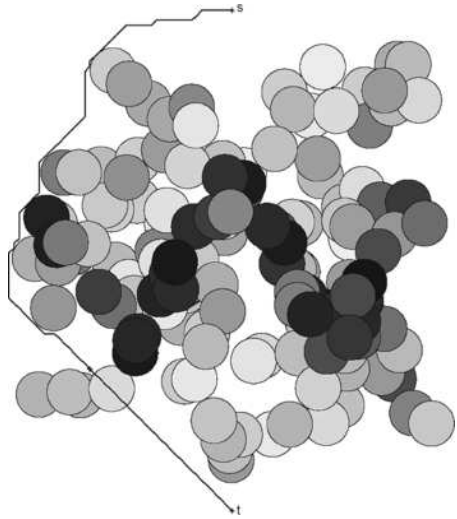}
\\
\footnotesize{(b) Navigation in CSR clutter${} + {}$W70:40}
\end{tabular}
\caption{The obstacle fields in Figures \protect\ref{figobst}\textup
{(b)} and \protect\ref{figobst}\textup{(c)}
as seen by the NAVA and the NAVA's navigation in these
fields.
Gray scale of disks [indicated in \textup{(a)} as intensity scale]
reflects $\rho$ of each disk, with darker colors indicating a higher
$\rho$.} \label{fignav}
\end{figure}

Shown in Figure~\ref{fignav} is how the NAVA sees the obstacle
fields illustrated in Figures~\ref{figobst}(b) and~\ref{figobst}(c),
respectively, and the $s-t$ walks taken by the NAVA as dictated by
the ARD algorithm. In Figure~\ref{fignav}(a), the NAVA performs a
total of 18 disambiguations and the total traversal length
(including the cost of disambiguations) is 311.8 units. This
particular walk turns out to be rather unfavorable (from the NAVA's
perspective), as the zero-risk $s-t$ walk length avoiding all the
disks has a traversal length of merely 151.3 units.
Such unfavorable
traversals occasionally happen, as the goal of the NAVA is to
minimize the \textit{expected} traversal length, and the actual walk
traversed can be much longer than the zero-risk $s-t$ walk based on
the outcomes of the disambiguations performed. In
Figure~\ref{fignav}(b), on the other hand, the NAVA performs only 1
disambiguation and the total traversal length is 152.2 units.

%

\section{Experimental setup and the statistical analysis}
\label{secanalysis}
Our particular experimental setup leads to a three-way ANOVA
problem. The treatment factors are the background clutter type, the
obstacle placement window and the number of obstacles.
The response variable is the NAVA's total traversal length from $s$ to $t$.
The first treatment factor has 6 levels, the second has 19, and the
third has 5 levels, resulting in a total of 570 treatment combinations.
Our primary goal here is to investigate whether there are any
(statistically significant) differences between traversal lengths of
different obstacle placement
windows for a given number of obstacles and a given clutter type.

For each one of these 570 treatment combinations, we ran 100 Monte
Carlo simulations. Each simulation consists of generating the
obstacle field (i.e., the obstacle pattern superimposed on a clutter
pattern) and executing the ARD algorithm to find the shortest $s-t$
walk. The runtime per simulation averaged over the 57,000
simulations was 9.5 seconds on a personal computer with an Intel
Core i7 processor with 2.8 gigahertz clock speed.

As discussed earlier, each background clutter realization is sampled to
have exactly 100 clutter disks via rejection sampling.
In order to exclude the source of variability due to
different clutter realizations for a given clutter type, we adopted a
repeated measures approach in our experiments. That is, we sampled only 100
clutter realizations from each one of the 6 clutter types
corresponding to each one of the 100 simulations for a given
treatment combination. Thus, a total of 600 clutter patterns were
generated for our experiments. For instance, the same CSR clutter realization
was used for all of the 95 obstacle pattern-obstacle number
combinations (19 obstacle patterns and 5 obstacle number levels) for
the first Monte Carlo simulation. For the second Monte Carlo
simulation, a different CSR realization was sampled and this
realization was
used for all of the 95 obstacle pattern-obstacle number
combinations and so on.

\subsection{Repeated measures ANOVA}
\label{secmethods}
The background clutter types
(abbreviations presented in parentheses) we consider
are Complete Spatial Randomness (CSR), inhomogeneous Poisson (IP) distribution,
Mat\'{e}rn (M) distribution, Thomas (T) distribution, hardcore (HC)
distribution and Strauss (S) distribution.
For convenience in
presentation, the obstacle types are sometimes numbered from 1 to
19, or labeled in a more descriptive fashion such as V90 which
stands for V-shaped obstacle window whose top left corner
$y$-coordinate is 90.
The 19 obstacle placement window types are
sampled within 4 different polygon shapes (a short notation is provided
in parentheses):
the entire $P$ window (P),
linear windows (L), V-shaped windows (V) and W-shaped windows (W).
The obstacle window numbering of 1 to 19 corresponds to P, L90, L80$, \ldots,$L20, V90, V80$, \ldots, $V50, W90, W80$, \ldots, $W50, respectively.
The obstacle number levels are $20,30,\ldots,60$. As mentioned
earlier, for precision in our analysis, we used the same background
clutter realization for each of the 95 obstacle types and
obstacle number combinations. Thus, we denote the traversal length as
$T_{ijkl}$, which is the traversal length of the measurement $l$ for
clutter type $i$, obstacle window type $j$, obstacle
number level $k$ with $l=1,2,\ldots,100$, $i=1,2,\ldots,6$,
$j=1,2,\ldots,19$, and $k=1,2,\ldots,5$, respectively. Clutter types
$1,2,\ldots,6$ correspond to CSR, IP, M, T, HC and S patterns,
respectively, and obstacle number levels $1,2,\ldots,5$ correspond
to $20, 30, \ldots,60$, respectively. Note that $T_{ijkl}$,
$T_{ij'kl}$, $T_{ijk'l}$, $T_{ij'k'l}$ are measured on the same
realization of the clutter type $i$, hence, these measures
are potentially correlated. In particular, the measurements on
consecutive obstacle number levels (with other factors being the same)
would be highly (and perhaps positively) correlated.
A similar trend can be expected for measurements within each type of
obstacle window
type (such as linear, V-shaped or W-shaped obstacle forms) as a
function of
the distance to the target (i.e., as a function of the magnitude of the
$y$ coordinate).
To take such correlation structure into account, we use \textit{repeated
measures ANOVA} techniques in our analysis to compare the traversal
length differences between treatment factors, and possibly existence
or lack of any interaction between these factors. Traditionally, repeated
measures ANOVA is employed when the measurements are taken on the
same subject over time [\citet{kuehl2000}], but here we are in a
similar but nontemporal situation. In our setup, each subject (i.e.,
background clutter realization) receives all of the 95 treatments
(clutter type, obstacle window and obstacle number
combinations). Besides, we do not need to randomize the order of the
treatments here, since when each treatment combination is applied
(i.e., each obstacle pattern is superimposed on the particular clutter
realization), we remove the previous data points that come from the
other factors. Hence, there is no carry-over effect of the
treatments in our study.


The assumptions of repeated measures ANOVA are similar to the standard
set of assumptions associated with usual ANOVA,
except that independence is not required and an assumption about the
relations among the repeated measures (sphericity) is added.
The repeated measures ANOVA assumptions are
(i) the dependent variable is normally distributed,
(ii) homogeneity of covariance matrices,
(iii) independence between predictor factors, and
(iv) sphericity, which means that the variances of the repeated
measures are all equal, and the correlations among the repeated
measures are all
equal
[\citet{tabachnick2006}, \citet{howell2010}].

Repeated measures ANOVA is robust to violations of the first two
assumptions [\citet{tabachnick2006}].
Besides,
the kernel density plots of the residuals (not presented) resemble that
of a Gaussian distribution.
(iii) is satisfied by construction in our experimental setup
(i.e., the factors clutter type, obstacle type and number of obstacle
levels are not dependent).
The violation of sphericity is the reason
we try various competing variance--covariance structures in addition to
compound symmetry
(to capture the dependence structure between repeated measures as much
as possible).
The main benefit of repeated measures ANOVA compared to usual ANOVA is
that with repeated measures ANOVA
we gain more precision in our results (i.e., the tests are more
powerful with more significant $p$-values).
Just as using paired differences in the two sample case compared to the
two independent samples case increases precision,
using repeated measures setup increases the precision compared to usual ANOVA.

In fact, we performed a pilot study to appraise the relative merit of
repeated measures ANOVA to usual ANOVA.
For the CSR clutter distribution, we generated 5 obstacle number levels
(of 20, 30$, \ldots,$60), where obstacle patterns also follow CSR
within the sampling window.
For the repeated measures ANOVA (called setup~I),
we used the same CSR clutter realization for each of $20, 30,\ldots,60$
obstacles,
and we repeated the procedure 100 times (i.e., in total there are 100
different CSR clutter realizations).
On the other hand,
for the usual ANOVA (called setup~II),
we used different CSR clutter realizations for each of $20, 30, \ldots,
60$ obstacles,
and we repeated the procedure 100 times (i.e., in total there are 500
different CSR clutter realizations).
When the data from setup I was analyzed with repeated measures ANOVA,
the obstacle number level was significant
(i.e., mean traversal lengths are different for the obstacle number
levels) with $F_{4,495}=45.91$,
where $F_{4,495}$ stands for $F$ test statistic with degrees of freedom
for numerator and denominator
being 4 and 495, respectively;
on the other hand, when the data from setup II was analyzed with usual
ANOVA, the obstacle number level was significant
with $F_{4,495}=27.75$.
Although the obstacle number is a significant factor in both cases,
the repeated measures setup yields a higher level of significance
(i.e., more precision and power) compared to the usual ANOVA setup.
A similar trend is observed for other clutter and obstacle type levels
(not presented), hence, the preference of the current setting over a
simple Monte Carlo setup.

In repeated measures ANOVA, we consider (some of or some variants of
the) four types of variance--covariance (var--cov) structure:
unstructured (UN), autoregressive (AR1), autoregressive heterogeneous
(ARH1) and
compound symmetry (CS).
The CS structure assumes a single
variance $\sigma^2$ for all treatment combinations and a single
covariance $\sigma_1$ for each pair of treatment combinations
(i.e., sphericity).
The CS var--cov structure in our setup with $\sigma_1=0$ implies the
usual 3-way ANOVA.
The UN variance--covariance structure assumes that
each variance and covariance is unique, that is, measurements in each
of the treatment combinations have a unique variance $\sigma_i^2$, and
each pair of treatment combinations has a unique covariance
$\sigma_{ij}$.
The AR1 var--cov structure assumes that
observations which are close (in some sense) are more correlated
than measures that are more distant.
For example, the measurements
on obstacle numbers 20 and 30 are more correlated than the
measurements on obstacle numbers 20 and 60 (with other factors being
the same).
So there is a single variance $\sigma^2$ for all 95 treatment
combinations and covariance $\sigma\times\rho^k$ where $k$ stands
for the order of the measurement.
In ARH1 var--cov structure, the variances are also different for the
treatment combination levels. So there is a unique variance
$\sigma_i^2$ for each treatment combination, and the covariance
structure is as in the autoregressive case [\citet{pinheiro2000}].
In our experimental design, it is also possible to further detail
the var--cov structure as autoregressive var--cov heterogeneous within
only the obstacle factor levels, or heterogeneous within all
treatment combinations, or within only the obstacle forms and so on.
We also employ Mauchly's sphericity test to determine the
appropriateness of CS var--cov structure in the repeated measures ANOVA
[\citet{kuehl2000}].
That is, we can assume the CS structure
only when the Mauchly's test yields an insignificant $p$-value.
In our comparison of the models with various var--cov structures, we apply
the model selection criteria as Akaike Information Criteria (AIC)
and also perform a $\chi^2$ test on the log likelihood function
[\citet{burnham2003}].

In what follows, the first section compares overall traversal
lengths for the three treatment factors. The next three sections (i.e.,
Sections~\ref{secoverall-length}--\ref{secobstacle-number})
present statistical comparison of traversal lengths at clutter type,
obstacle form and obstacle number level factors,
respectively. The last section gives statistically best performing
obstacle type and number combinations at each clutter type.

\subsection{Overall comparison of traversal lengths}
\label{secoverall-length}
\begin{figure}
\centering
\begin{tabular}{@{}cc@{}}

\includegraphics{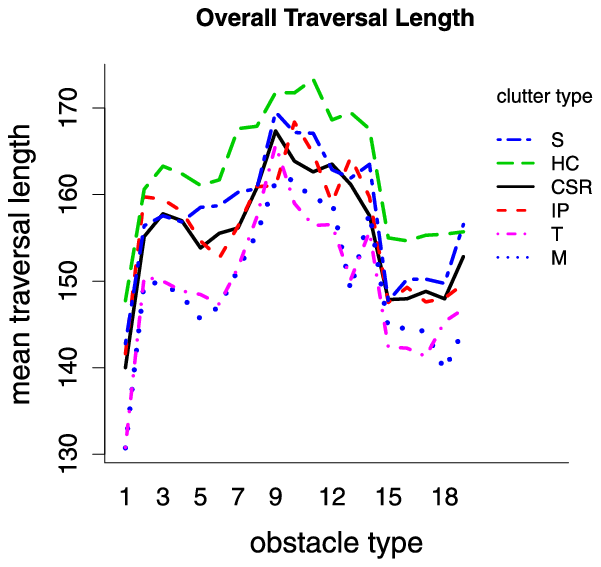}
 & \includegraphics{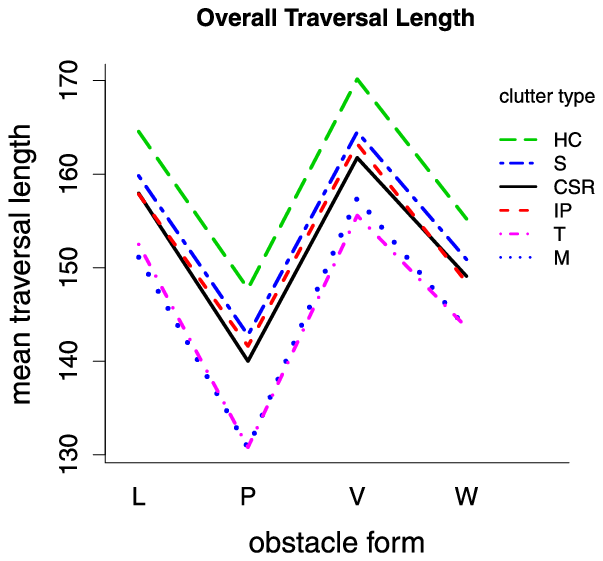}\\[-3pt]
\footnotesize{(a)} & \footnotesize{(b)}\\[6pt]

\includegraphics{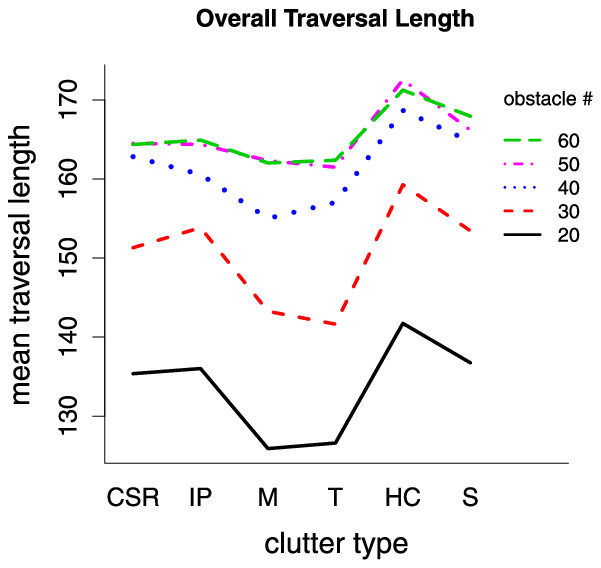}
 & \includegraphics{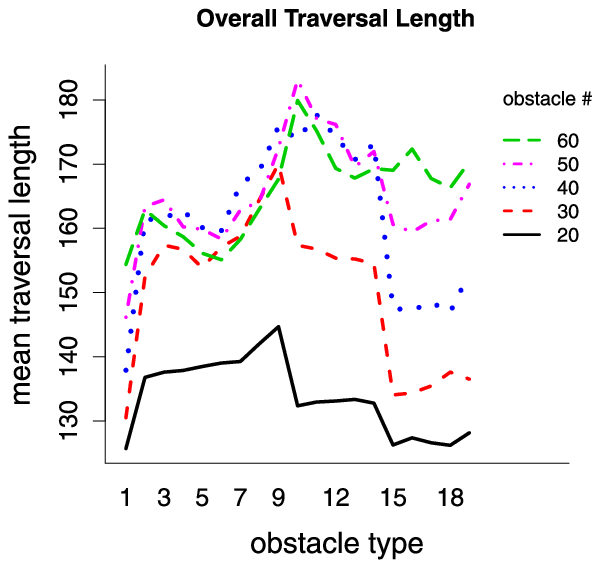}\\[-3pt]
\footnotesize{(c)} & \footnotesize{(d)}\vspace*{3pt}
\end{tabular}
\centering
\begin{tabular}{@{}c@{}}

\includegraphics{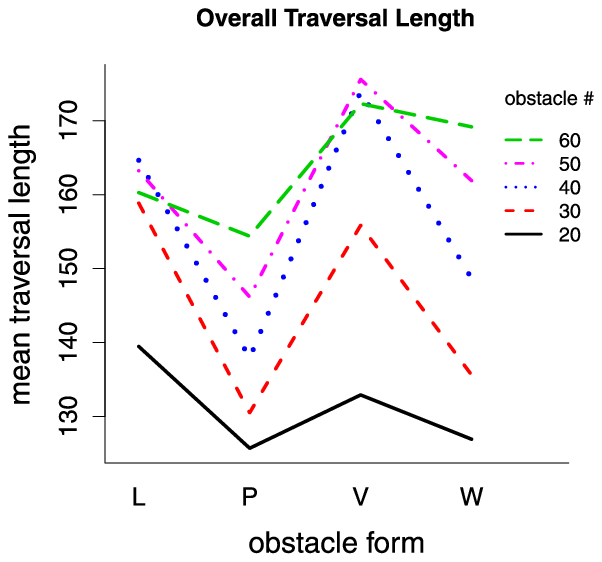}
\\[-3pt]
\footnotesize{(e)}
\end{tabular}
\caption{The profile plots for each pair of treatment factors
(obstacle type/form, clutter type and obstacle number)
when the other factor is ignored.}\label{figoverall-interaction}
\end{figure}

We first investigate the types and levels
of interaction for each pair of treatment combination factors.
The profile (or interaction) plots are shown in Figure \ref
{figoverall-interaction}.
We also test for interaction between each pair of treatment factors.
When obstacle number levels are ignored (i.e., when only interaction
between obstacle types/forms and clutter types are
considered), we find that obstacle and clutter types do not have
significant interaction ($p=0.5258$), neither do obstacle forms and
clutter types ($p=0.4811$), which means the trends in mean lengths
plotted in Figures~\ref{figoverall-interaction}(a) and (b) are not
significantly different from being parallel. Hence, it is reasonable
to compare the mean traversal lengths for clutter and obstacle types
(i.e., the main effects of clutter and obstacle types), and we find
that travel lengths are significantly different between background
clutter types ($p<0.0001$) and between obstacle types
($p<0.0001$). Likewise, traversal lengths are significantly
different between clutter types ($p<0.0001$) and between
obstacle forms ($p<0.0001$). Notice also that, on the average at
each obstacle type or form, Mat\'{e}rn and Thomas (i.e., clustered)
clutter types tend to yield shorter traversal lengths, while
hardcore and Strauss (i.e., regular) clutter types tend to
yield longer traversal lengths. Hardcore clutter tends to yield the
longest traversal lengths, which suggests that the more regular the
clutter type, the longer the traversal lengths. Furthermore,
at each obstacle type or form, the average traversal lengths (in
ascending order) are for P, W-shaped, linear and V-shaped obstacle
forms.

%

When clutter types are ignored (i.e., when only interaction between
obstacle types/forms and obstacle number levels are considered), we
find significant interaction between obstacle type and obstacle
number levels ($p<0.0001$), and between obstacle form and obstacle
number levels ($p<0.0001$), which means the trends in mean lengths
plotted in Figures~\ref{figoverall-interaction}(d) and (e) are
significantly nonparallel. Hence, it is not reasonable to compare
the mean traversal lengths for obstacle types/forms and obstacle
number levels, but instead, for example, it will make sense to
compare the mean length values for obstacle number levels at each
obstacle type or form. At P and W-shaped obstacle forms, traversal
lengths tend to increase as the obstacle number increases; at linear
and V-shaped obstacle forms, traversal lengths exhibit a concave-down trend
(i.e., increase, reach a peak and then decrease);
for the linear and V-shaped windows the shortest lengths
occur at 20 obstacles, and longest lengths occur at 40 obstacles.
We believe that the concave-down trend is due to the increase in the
disk (obstacle and clutter) density that makes the NAVA decide to
traverse along the boundary more often, which reduces the traversal
length, since the NAVA avoids the disambiguation costs.
Hence, a~similar concave-down trend (with larger obstacle
numbers) is expected to occur for P and W-shaped obstacle forms as well.
Moreover, for 20 and 30 obstacles, the highest (average)
traversal lengths occur for linear obstacle forms, and for 40--60
obstacles, longest traversal lengths occur for V-shaped obstacle
forms. At each obstacle number level, the shortest traversal lengths
occur for the P obstacle form.

When obstacle types are ignored (i.e.,
when only interaction between clutter type and obstacle number levels
are considered),
we find significant interaction between clutter type and obstacle
number levels ($p<0.0001$),
which means the trend in mean length plotted in Figure \ref
{figoverall-interaction}(c)
is significantly nonparallel.
Hence, we compare the mean length values for
obstacle number levels at each clutter type. On the average at each
clutter type, traversal lengths tend to increase as the obstacle
number increases (up to 50 obstacles), but the lengths for 50 and 60
obstacles are very similar. At each obstacle number level, the
longest traversal lengths occur for hardcore clutter type,
and the shortest traversal lengths occur for Mat\'{e}rn and Thomas
clutter types.

The shortest and longest traversal length performances (i.e., the
worst and best performances from the OPA perspective) are presented
in Table~\ref{tabshortest-highest-lengths}. In our overall
comparison, the shortest length is about 116 units which occurs at
T:W60:20, T:P:20 and M:P:20 treatment combinations, and the
longest length is about 190 which occurs at HC:V90:50 treatment
combination. Our initial (overall) interaction analysis suggests
that it is more reasonable to compare the lengths for each pair of
treatment factors at specific levels of the other factor
different from both factors in the pair.

In Sections~\ref{secbg-types}--\ref{secobstacle-number},
the profile plots,
model comparison tables
and their detailed discussions
are deferred to the technical report \citet{aksakalliTR-OOP-2012}.

%
\begin{table}
\tabcolsep=0pt
\caption{The shortest and longest traversal lengths and the
corresponding treatment types for
overall comparisons, and comparisons at specific clutter types,
obstacle forms and obstacle numbers}\label{tabshortest-highest-lengths}
\begin{tabular*}{\textwidth}{@{\extracolsep{\fill}}lcccc@{}}
\hline
& \multicolumn{2}{c}{\textbf{Shortest}} &
\multicolumn{2}{c@{}}{\textbf{Longest}}\\[-4pt]
& \multicolumn{2}{c}{\hrulefill} & \multicolumn{2}{c@{}}{\hrulefill}\\
& \textbf{Traversal} & \textbf{Treatment} & \textbf{Traversal} & \textbf
{Treatment}\\
& \textbf{length (s)} & \textbf{type (s)} & \textbf{length (s)} &
\textbf{type (s)}\\
\hline
 \multicolumn{5}{c}{Overall} \\
& 116.16, 116.47, & T:W60:20, T:P:20, & 190.29 & HC:V90:50\\
& 116.77 & M:P:20 & & \\[6pt]
\multicolumn{5}{c}{Clutter type} \\
CSR & 127.56, 127.76 & P:20, W60:20 & 178.68, 178.98, & V70:50,
V50:40,\\
& & & 179.12 & V80:50\\
Inhom. & 126.87 & W90:20 & 188.96 & V90:60\\
\quad Poisson & & & & \\
Mat\'{e}rn & 116.77 & P:20 & 184.39 & V90:50\\
Thomas & 116.16, 116.47 & W60:20, P:20 & 181.48 & V90:60\\
Hardcore & 134.42, 134.85, & W80:20, W90:20, & 190.29 & V90:50\\
& 135.05, 135.50, & W70:20, W50:20, & & \\
& 135.81, 135.95 & P:20, W60:20 & & \\
Strauss & 128.47 & P:20 & 188.35, 188.77 & V90:50, V80:40\\[6pt]
\multicolumn{5}{c}{Obstacle form} \\
CSR & 116.47, 116.77 & T:20, M:20 & 163.86 & HC:60 \\
Linear & 127.32 & L40:M:20 & 184.42 & L20:HC:40\\
V-Shaped & 119.23 & V50:T:20 & 190.29 & V90:HC:50\\
W-Shaped & 116.16 & W60:T:20 & 176.91, 177.18, & W50:HC:60, W80:HC:60,
\\
& & & 177.36, 177.56, & W50:S:60, W60:T:60,\\
& & & 177.76 & W80:S:60\\[6pt]
\multicolumn{5}{c}{Number of obstacles} \\
20 & 116.16, 116.47, & T:W60, T:P, & 156.27 & HC:L40 \\
& 116.77 & M:P & & \\
30 & 121.73, 122.25 & T:P, M:P & 179.43 & HC:L20\\
40 & 128.95 & M:P & 187.57, 188.77 & HC:V80, S:V80\\
50 & 135.04 & M:P & 190.29 & HC:V90 \\
60 & 143.09 & T:P & 188.96 & IP:V90 \\
\hline
\end{tabular*}
\end{table}
%

\subsection{Analysis of traversal lengths at each background clutter type}
\label{secbg-types}
We investigate and test the interaction between
obstacle type/form and obstacle number at each background clutter type.
At each background clutter type,
we find significant interaction between obstacle form and obstacle
number levels ($p<0.0001$ for each)
and between obstacle type and obstacle number levels ($p<0.0001$ for each),
hence, we do not test for the main effects of obstacle types/forms and
obstacle number levels. At each background clutter type,
on the average at P and W-shaped
obstacle forms, traversal lengths tend to increase as the obstacle
number increases; at linear and V-shaped obstacle forms, traversal
lengths exhibit a concave-down trend:
The longest lengths occur at 40 obstacles for each obstacle form at CSR
and Strauss clutter, and for linear obstacles at inhomogeneous Poisson,
Thomas and hardcore clutters and at 50 or 60 obstacles for V-shaped
obstacles at inhomogeneous Poisson, Thomas and hardcore clutters.
The shortest lengths occur at 20 obstacles.

The shortest and longest
traversal lengths together with the corresponding treatment
combinations are presented in Table
\ref{tabshortest-highest-lengths}.
Presented below is further discussion
on traversal lengths for each clutter type.\vspace*{6pt}

 \textit{CSR clutter}:
The shortest traversal length is about 127 which occurs at P:20, W60:20
treatment types, and the longest traversal length is about
179 which occurs at V70:50, V50:40, V80:50 treatment types.
Moreover, for 20 and 30 obstacles, the longest traversal
lengths occur for linear obstacle forms, and for 40--60 obstacles,
longest traversal lengths occur for V-shaped obstacle forms. At each
obstacle number level, the shortest traversal lengths occur for the
P obstacle form.

 \textit{Inhomogeneous Poisson clutter}:
The shortest traversal length is about 126 which occurs at W90:20
treatment type, and the longest traversal length is about 188 which
occurs at V90:60 treatment type.
For 20 obstacles, shortest traversal
lengths occur at W-shaped obstacle forms, and the longest traversal
lengths occur for linear obstacle forms; and for 30--60 obstacles,
the shortest traversal lengths occur for the P obstacle form and
longest traversal lengths occur for V-shaped obstacle forms.

 \textit{Mat\'{e}rn clutter}:
The shortest traversal length is about 117 which occurs at P:20
treatment type, and the longest traversal length is about 184 which
occurs at V90:50 treatment type.
For 20 and 30 obstacles, the longest
traversal lengths occur for linear obstacle forms, and for 40--60
obstacles, longest traversal lengths occur for V-shaped obstacle
forms. At each obstacle number level, the shortest traversal lengths
occur for the P obstacle form.

 \textit{Thomas clutter}:
The shortest traversal length is about 116 which occurs at W60:20, P:20
treatment types, and the longest traversal length is about 181
which occurs at V90:60 treatment type.
For 20 and 30 obstacles, the longest traversal lengths
occur for linear obstacle forms, and for 40--60 obstacles, longest
traversal lengths occur for V-shaped obstacle forms. At each
obstacle number level, the shortest traversal lengths occur for the
P obstacle form.

 \textit{Hardcore clutter}:
The shortest traversal length is about 135 which occurs at W80:20,
W90:20, W70:20, W50:20, P:20, W60:20 treatment types, and the
longest traversal length is about 190 which occurs at V90:50
treatment type.
For 20 obstacles,
shortest traversal lengths occur at the P obstacle form, and the
longest traversal lengths occur for linear obstacle forms; for 30
obstacles, shortest traversal lengths occur at W-shaped obstacle
types, and the longest traversal lengths occur for linear obstacle
forms; and for 40--60 obstacles, the shortest traversal lengths occur
for the P obstacle form and longest traversal lengths occur for
V-shaped obstacle forms.

 \textit{Strauss clutter}:
The shortest traversal length is about 128 which occurs at P:20
treatment type, and the longest traversal length is about 188 which
occurs at V90:50, V80:40 treatment types.
For 20 and 30 obstacles, the longest
traversal lengths occur for linear obstacle forms, and for 40--60
obstacles, longest traversal lengths occur for V-shaped obstacle
forms. At each obstacle number level, the shortest traversal lengths
occur for the P obstacle form.

\subsection{Analysis of traversal lengths at each obstacle form}
\label{secobs-forms}
We investigate the pairwise interaction between
background clutter type, obstacle type and obstacle number at each
obstacle form.
Note that only clutter type and obstacle number interaction is well
defined for the P obstacle form.
For other obstacle forms each pair of interaction is possible.
Our statistical analysis results are given below.\vspace*{6pt}

 \textit{The P obstacle form}:
We find no significant interaction between clutter type and obstacle
number levels ($p=0.5699$). Hence, we test for
main effects of clutter types and obstacle number levels. The
traversal lengths are significantly different between background
clutter types ($p<0.0001$) and between obstacle number levels
($p<0.0001$).

 \textit{Linear obstacle form}:
We find significant interaction between clutter type and obstacle
number levels ($p=0.0009$), between obstacle type and obstacle
number levels ($p<0.0001$), and between obstacle type and clutter
type ($p<0.0001$).
So, it is not reasonable to test for
the main effects of obstacle types, clutter types or obstacle
number levels.

 \textit{V-shaped obstacle form}:
We find significant interaction between clutter type and obstacle
number levels ($p=0.0004$), and between obstacle type and obstacle
number levels ($p=0.0142$).
So, it is not reasonable to compare the main effects
of clutter types and obstacle number levels nor the main effects of
obstacle types and obstacle number levels. On the other hand, there
is no significant interaction between obstacle type and clutter type
($p=0.2526$). So, we
compare the main effects of obstacle types and clutter types. The
traversal lengths are significantly different between background
clutter types ($p<0.0001$) and between obstacle types ($p<0.0001$).

 \textit{W-shaped obstacle form}:
We find significant interaction between clutter type and obstacle
number levels ($p<0.0001$).
So, it is not reasonable to compare
the main effects of clutter type and obstacle number levels here.
But we find no significant interaction between obstacle type and
obstacle number levels ($p=0.1298$), and between obstacle type and
clutter type ($p=0.6028$).
So, we compare the
main effects of obstacle types and obstacle numbers and to compare
for obstacle and clutter types. The traversal lengths are
significantly different between obstacle types ($p=0.0011$) and
between obstacle number levels ($p<0.0001$) ignoring clutter types,
and traversal lengths are significantly different between clutter
types ($p<0.0001$) and between obstacle types ($p=0.0038$) ignoring
obstacle number levels.\vspace*{6pt}

Analysis of traversal lengths for each obstacle form is given below.

 \textit{The P obstacle form}:
The shortest length is about 116.5 which occurs at T:20, M:20 treatment
types, and the longest length is about 164
which occurs at HC:60 treatment type. On the average at each
clutter type, traversal lengths tend to increase as the obstacle
number increases. For 20, 30 and 60 obstacles, the shortest
traversal lengths occur for the Thomas clutter type, and for 40 and 50
obstacles, shortest traversal lengths occur for the Mat\'{e}rn clutter
type. At each obstacle number level, the longest traversal lengths
occur for the hardcore clutter type. Therefore, for the P obstacle form,
traversal lengths tend to be shorter for clustered clutter types, and
longer for regular clutter types.

 \textit{Linear obstacle form}:
The shortest length is about 127 which occurs at L40:M:20
treatment type, and the longest length is about 184 which
occurs at L20:HC:40 treatment type. On the average at each
clutter type, traversal lengths exhibit a concave-down trend as
obstacle number increases (for Mat\'{e}rn clutter, the
longest length occurs at 50 obstacles, and for other clutter types,
the longest lengths occur at 40 obstacles; for each clutter type
shortest lengths occur at 20 obstacles). For 30 obstacles, the
shortest traversal lengths occur for the Thomas clutter type, and for
other obstacle number levels, shortest traversal lengths occur for the
Mat\'{e}rn clutter type. At each obstacle number level, the longest
traversal lengths occur for the hardcore clutter type. Therefore,
for linear obstacle form, traversal lengths tend to be shorter for
clustered clutter types, and longer for regular clutter types.
As obstacle type level increases (from 2 to 9),
length tends to increase as well. That is, as the distance of the
linear window to the coast (where $t$ is located) increases, so does
the traversal length.
At L90 and L80, longest length occurs at 50 obstacles, at other
linear windows, the longest lengths occur at 40 obstacles. At each
obstacle type, the shortest lengths occur at 20 obstacles. At the
Strauss clutter type, length increases as obstacle number increases,
at CSR and hardcore clutter types, length increases, decreases to a
(local) minimum and then increases again as obstacle number
increases, and at other clutter types, length tends to decrease to a
minimum and then increases as obstacle number increases. At Strauss
and hardcore clutters, shortest length occurs at L90 and at other
clutter types shortest lengths occur at L50 or L60. At each clutter
type, longest lengths occur at L20 (i.e., at linear window closest
to the coast).

 \textit{V-shaped obstacle form}:
The shortest length is about 119 which occurs at V50:T:20
treatment type, and the longest length is about 190 which
occurs at V90:HC:50 treatment type. On the average at each
clutter type, traversal lengths exhibit a concave-down trend as
obstacle number increases (for CSR and Strauss clutter
types, the longest length occurs at 40 obstacles, and for other
clutter types, the longest lengths occur at 50 obstacles; for each
clutter type shortest lengths occur at 20 obstacles). For 40
obstacles, the shortest traversal lengths occur for the Mat\'{e}rn
clutter type, and for other obstacle number levels, shortest
traversal lengths occur for the Thomas clutter type. At each obstacle
number level, the longest traversal lengths occur for the hardcore
clutter type. Therefore, for V-shaped obstacle form, traversal
lengths tend to be shorter for clustered clutter types, and
longer for regular clutter types. For 20 and 30 obstacles,
length trend tends to be flat (i.e., not changing) as obstacle type
level increases (from 10 to 14); for 40 obstacles, length tends to
exhibit a concave-down trend as obstacle type level
increases; and for 50 and 60 obstacles, length tends to decrease as
obstacle type level increases. That is, for large obstacle numbers,
length tends to decrease as the distance of the V-shaped window to
the coast decreases. At V90 and V70 windows, longest lengths occur
at 50 obstacles, while at other windows, longest lengths occur at 40
obstacles. At each V-shaped obstacle type, the shortest lengths
occur at 20 obstacles. The length trend is similar at clustered
clutter types (hardcore and Strauss), and likewise at regular
clutter types (Mat\'{e}rn and Thomas). For clustered clutters, the
longest lengths occur at V80 window; and the shortest lengths occur
at V50 for the hardcore clutter, and at V60 for the Strauss clutter. For
regular clutters, longest lengths occur at V90 window and shortest
occurs at V60 window. For the inhomogeneous Poisson clutter, longest
length occurs at V90 window, and shortest occurs at V70 window. For the
CSR clutter, longest length occurs at V90--V70 windows, and shortest
occurs at V50 window.

 \textit{W-shaped obstacle form}:
The shortest length is about 116 which occurs at W60:T:20
treatment type, and the longest length is about 177 which
occurs at W50:HC:60, W80:HC:60, W50:S:60, W60:T:60, W80:S:60
treatment types. On the average at each clutter type, traversal
lengths tend to increase as obstacle number increases. For 20, 40
and 50 obstacles, the shortest traversal lengths occur for the Thomas
clutter type; for 30 obstacles shortest traversal length occurs for the
Mat\'{e}rn clutter type; and for 60 obstacles, shortest traversal
length occurs for the inhomogeneous Poisson clutter type. At each
obstacle number level, the longest traversal lengths occur for the
hardcore clutter type. Therefore, for W-shaped obstacle form,
traversal lengths tend to be shorter for clustered clutter types, and
longer for regular clutter types. At each
obstacle number level, the length trend tends to be flat as obstacle
type level increases (from 15 to 19). That is, length seems not to
depend strongly on the distance of the W-shaped window to the $x$-axis.
For the
hardcore clutter, length tends to be flat as obstacle type level
increases (from 15 to 19). For the CSR and Strauss clutter type, the
length trend is similar. For the CSR clutter, the shortest length occurs
at W90, W80 and W60 windows; for the Strauss clutter at W90 window, for the
inhomogeneous Poisson clutter at W90, W70 and W60 windows, for the
Thomas clutter at W90, W80 and W70 windows, and for the Mat\'{e}rn
clutter at W60 window. At each clutter type, the longest length
occurs at W50 window (i.e., when the obstacles are closest to the
coast).

\subsection{Analysis of traversal lengths at each obstacle number level}
\label{secobstacle-number}
At obstacle number levels of 20, 50 and 60,
we find significant interaction between obstacle type and background
clutter type
($p=0.0379$, $0.0042$ and $0.0006$, resp.).
Hence, it is not reasonable to compare the main effects of obstacle
types and clutter types.
At obstacle number levels of 30 and 40,
we find no significant interaction between obstacle type and background
clutter type
($p=0.3592$ and $0.9340$, resp.).
Hence,
we test for the main effects of obstacle and clutter types.
The traversal lengths are significantly different between background
clutter types ($p<0.0001$)
and between obstacle types ($p<0.0001$).

At the obstacle number level of 60,
we find significant interaction between obstacle form and background
clutter type ($p=0.0124$).
Hence, it is not reasonable to compare the main effects of obstacle
types and clutter types.
At obstacle number levels of 20, 30, 40 and 50,
we find no significant interaction between obstacle form and background
clutter type
($p=0.2207$, $0.6824$, $0.8876$ and $0.0895$, resp.).
Hence, it is reasonable to compare the main effects of obstacle forms
and clutter types.
The traversal lengths are significantly different between background
clutter types ($p<0.0001$)
and between obstacle forms ($p<0.0001$).\vspace*{6pt}

 \textit{20 obstacles}:
The shortest traversal length is about 116.5 which occurs at T:W60,
T:P, M:P
treatment types, and the longest length is about 156 which
occurs at HC:L40 treatment type. On the average at each obstacle
form, the longest traversal lengths occur for the hardcore clutter type,
and the shortest traversal lengths occur for the Mat\'{e}rn and Thomas
clutter types. The traversal lengths for CSR, Strauss and
inhomogeneous Poisson clutters are similar, although the Strauss clutter
tends to have longer length values. The mean traversal lengths can
be sorted in ascending order as P, W-shaped, V-shaped and linear
obstacle forms at each clutter type.

 \textit{30 obstacles}:
The shortest length is about 121 which occurs at T:P, M:P
treatment types, and the longest length is about 179 which
occurs at HC:L20 treatment type. On the average at each obstacle
form, the longest traversal lengths occur for the hardcore clutter type,
and the shortest traversal lengths occur for Mat\'{e}rn and Thomas
clutter types. The traversal lengths for CSR and Strauss clutters
are similar, although Strauss clutter tends to have longer length
values. For the inhomogeneous Poisson clutter type, the mean traversal
lengths can be sorted in ascending order as P, W-shaped, linear and
V-shaped forms; for other clutter types, the mean traversal lengths
can be sorted in ascending order as P, W-shaped, V-shaped and
linear obstacle forms.

 \textit{40 obstacles}:
The shortest length is about 129 which occurs at M:P treatment
type, and the longest length is about 188 which occurs at HC:V80, S:V80
treatment types. On the average at P, linear and W-shaped
obstacle forms, the shortest traversal lengths occur for Mat\'{e}rn
and Thomas clutter types, and at V-shaped obstacle types, the
shortest traversal lengths occur for Mat\'{e}rn, Thomas and
inhomogeneous Poisson clutter types. At each obstacle type, the
longest traversal lengths occur for the hardcore clutter type. The
traversal lengths for CSR and Strauss clutters are similar, although the
Strauss clutter tends to have longer length values. For each clutter
type, the mean traversal lengths can be sorted in ascending order as
P, W-shaped, linear, and V-shaped forms.

 \textit{50 obstacles}:
The shortest length is about 135 which occurs at M:P treatment
type, and the longest length is about 190 which occurs at HC:V90
treatment type. On the average at P and linear obstacle forms, the
shortest traversal lengths occur for Mat\'{e}rn and Thomas clutter
types; at V-shaped obstacle types, the shortest traversal lengths
occur for Thomas and CSR clutter types; and at W-shaped obstacle
forms, the shortest traversal lengths occur for Thomas clutter type.
At each obstacle type, the longest traversal lengths occur for
hardcore clutter type. The traversal lengths for CSR, inhomogeneous
Poisson, and Strauss clutters are similar at P and linear obstacle
forms, although Strauss clutter tends to have longer length values;
and traversal lengths for Mat\'{e}rn, inhomogeneous Poisson, and
Strauss clutters are similar at V- and W-shaped obstacle forms. For
each clutter type, the mean traversal lengths can be sorted in
ascending order as P, W-shaped, linear and V-shaped forms.

 \textit{60 obstacles}:
The shortest length is about 143 which occurs at T:P treatment
type, and the longest length is about 189 which occurs at IP:V90
treatment type. On the average at each obstacle form, the longest
traversal lengths occur for the hardcore clutter type and then for the
Strauss clutter type. At P and V-shaped obstacle forms, the shortest
traversal lengths occur for Thomas clutter types; at the linear obstacle
form, the shortest traversal lengths occur for Thomas and Mat\'{e}rn
clutter types; and at W-shaped obstacle forms, the shortest
traversal lengths occur for the inhomogeneous Poisson clutter type. The
traversal lengths for CSR and inhomogeneous Poisson clutters are
similar at P and linear obstacle forms; traversal lengths for CSR
and Mat\'{e}rn clutters are similar at V-shaped obstacle forms; and
traversal lengths for Thomas and Strauss clutters are similar at
W-shaped obstacle forms. Furthermore, traversal lengths for Strauss
and inhomogeneous Poisson clutters are similar at V-shaped obstacle
forms. For each clutter type except Thomas clutter, the mean
traversal lengths can be sorted in ascending order as P, linear,
W-shaped and V-shaped forms; and for the Thomas clutter, the mean
traversal lengths can be sorted in ascending order as P, linear,
V-shaped and W-shaped forms.

\subsection{Comparison of best performers at each clutter type}
\label{seccomp-best-perf}
From the OPA's perspective,
it is more desirable to make the NAVA traverse longer lengths to reach the\vadjust{\goodbreak}
target point. Furthermore, in our scenario the OPA is assumed to
have no control on the clutter type, but can only determine/know
the clutter type (but not the actual locations of the clutter
disks). Hence, for a given background clutter type, it is desirable
to determine the obstacle type-obstacle number combination
that yields the longest traversal lengths.
This combination is referred to as ``best performer'' henceforth.
The overall best performer and best performers for each
obstacle type at each clutter type are presented in Table
\ref{tabbest-perform}.

%
\begin{table}
\tabcolsep=0pt
\caption{The best performers (i.e., the
obstacle type-obstacle number combination with the longest
traversal lengths) for each background clutter type and the
corresponding average traversal lengths. The best performer row is
labeled as ``trt comb.,'' and obstacle form as ``obs. form.'' The
``Overall'' column is the best performer treatment combination at
each clutter type. The obstacle type with the largest traversal
lengths (that are significantly larger than the rest at the 0.01 level)
are marked in bold face, and the traversal lengths that are not
significantly different at the 0.05 level but different at the 0.10 level
are marked with an asterisk *}\label{tabbest-perform}
\begin{tabular*}{\textwidth}{@{\extracolsep{\fill}}lccccc@{}}
\hline
\multicolumn{6}{c}{CSR clutter} \\
obs. form & Overall & P & \textbf{Linear} & \textbf{V-Shaped} &
W-Shaped*\\
trt comb. & V80:50 & P:60 & L20:50 & V80:50, V50:40, V70:50 & W50:60,
W50:50\\
mean length & -- & 154.36 & 175.93 & 178.93 & 172.22 \\[3pt]
\multicolumn{6}{c}{Inhomogeneous Poisson clutter} \\
obs. form & Overall & P & Linear & \textbf{V-Shaped} & W-Shaped\\
trt comb. & V90:60 & P:60 & L20:40 & V90:60 & W90:60\\
mean length & -- & 155.46 & 170.40 & 188.96 & 173.60 \\[3pt]
\multicolumn{6}{c}{Mat\'{e}rn clutter} \\
obs. form & Overall & P & Linear & \textbf{V-Shaped} & W-Shaped\\
trt comb. & V90:50 & P:60 & L20:40 & V90:50 & W90:60\\
mean length & -- & 150.59 & 173.23 & 184.39 & 171.21 \\[3pt]
\multicolumn{6}{c}{Thomas clutter} \\
obs. form & Overall & P & \textbf{Linear} & \textbf{V-Shaped} & \textbf
{W-Shaped}\\
trt comb. & V90:60 & P:60 & L20:50 & V90:60 & W60:60\\
mean length & -- & 143.09 & 176.15 & 181.48 & 177.56 \\[3pt]
\multicolumn{6}{c}{Hardcore clutter} \\
obs. form & Overall & P & Linear & \textbf{V-Shaped} & W-Shaped\\
trt comb. & V90:50 & P:60 & L20:40 & V90:50 & W80:60, W50:60\\
mean length & -- & 163.86 & 174.43 & 190.29 & 177.04 \\[3pt]
\multicolumn{6}{c}{Strauss clutter} \\
obs. form & Overall & P & Linear* & \textbf{V-Shaped} & W-Shaped\\
trt comb. & V80:40 & P:60 & L20:40 & V80:40, V90:50 & W80:60, W50:60\\
mean length & -- & 158.94 & 179.45 & 188.56 & 177.56 \\
\hline
\end{tabular*}
\end{table}

Since there are multiple
best performer obstacle type-obstacle number combinations at some
clutter types (see Table~\ref{tabbest-perform}),
we compare the traversal lengths of best performers for obstacle form
levels at each clutter type.
At each background clutter type,
we consider the following model with four different var--cov structures:
%
\begin{equation}
\label{eqnbest-perf-model}
T_{ij}=\mu_0+\mu^{\mathrm{OF}}_i+\varepsilon_{ij},
\end{equation}
where
$\mu_0$ is the overall mean,
$\mu^{\mathrm{OF}}_i$ is the main effect of obstacle form $i$,
and
$\varepsilon_{ij}$ is the error term
for $i=1,2,3,4$ (which correspond to P, linear, V-shaped and W-shaped
obstacle forms)
and $j=1,2,\ldots,n_i$,
where $n_i$ is $k\times100$ with $k$ being the number of treatment
combinations that are best performers.
For example, for the CSR clutter type, $k=1$ for the P obstacle type
and $k=3$ for the V-shaped obstacle type.
The var--cov structures we consider are
compound symmetry (CS),
unstructured (UN),
autoregressive var--cov structure (AR1),
autoregressive heterogeneous (ARH1).
When Mauchly's test is performed,
we obtain $p<0.0001$ for the CSR clutter,
$p=0.2172$ for the inhomogeneous Poisson clutter,
$p=0.1032$ for the Mat\'{e}rn clutter,
$p=0.0005$ for the Thomas clutter,
$p=0.0764$ for the hardcore clutter
and
$p=0.0002$ for the Strauss clutter.
That is, for the inhomogeneous Poisson clutter,
we can assume CS in var--cov structure,
and for Mat\'{e}rn and hardcore clutters, it is a close call for being
significant,
so we also consider the AIC values and likelihood ratio $p$-values
which are presented in Table~\ref{tabbest-perf-mod-comp}.
Notice that at the inhomogeneous Poisson clutter the model with CS
var--cov structure
(which agrees with the result of Mauchly's test)
and
at other clutter types, the model with ARH1 var--cov structure
seems to be the best model,
since these models have the smallest AIC values.
The $p$-values are based on the likelihood ratio of the model with
smallest AIC
and the model in the corresponding row.
Hence, when Mauchly's test yields an almost significant $p$-value,
we also consider the model selection criteria such as AIC and log
likelihood measures.
If the likelihood ratio test is not significant for two models,
we follow the common practice of picking the simpler model (i.e., the
model with fewer parameters).
With the best models,
we observe significant differences between obstacle types.

%
\begin{table}
\caption{The comparisons of the models for the best performer treatment
combinations
as in equation (\protect\ref{eqnbest-perf-model})
for each background clutter type.
The column labels are degrees of freedom (df),
Akaike information criterion (AIC),
negative log likelihood value,
likelihood ratio test ($L$-ratio).
The models are with compound symmetry (CS),
autoregressive (AR1),
autoregressive heterogeneous (ARH1)
and
unstructured (UN)
var--cov structure.
The likelihood ratio ($L$-ratio)
and the associated $p$-value are with respect to the model with
the smallest AIC value}\label{tabbest-perf-mod-comp}
\begin{tabular*}{\textwidth}{@{\extracolsep{\fill}}lcccd{3.2}d{2.4}@{}}
\hline
\textbf{var--cov structure} & \textbf{df} & \textbf{AIC} & \textbf{$\bolds{-}$log
likelihood} & \multicolumn{1}{c}{$\bolds{L}$\textbf{-ratio}} &
\multicolumn{1}{c@{}}{$\bolds{p}$\textbf{-value}}\\
\hline
\multicolumn{6}{c}{CSR clutter type} \\
CS & \phantom{0}6 & 7490.54 & 3739.27 & 101.98 & <\!0.0001\\
UN & 29 & 7421.18 & 3681.59 & 13.37 & 0.8608\\
AR1 & \phantom{0}6 & 7487.76 & 3737.88 & 99.21 & <\!0.0001\\
ARH1 & \phantom{0}9 & 7394.55 & 3688.28 & \multicolumn{1}{c}{--} &
\multicolumn{1}{c@{}}{--}\\[3pt]
\multicolumn{6}{c}{Inhomogeneous Poisson clutter type} \\
CS & \phantom{0}6 & 4048.91 & 2018.45 & 29.23 & <\!0.0001\\
UN & 14 & 4028.11 & 2000.05 & 7.57 & 0.1815\\
AR1 & \phantom{0}6 & 4053.76 & 2020.88 & 34.08 & <\!0.0001\\
ARH1 & \phantom{0}9 & 4025.68 & 2003.84 & \multicolumn{1}{c}{--} &
\multicolumn{1}{c@{}}{--}\\[3pt]
\multicolumn{6}{c}{Mat\'{e}rn clutter type} \\
CS & \phantom{0}6 & 4159.92 & 2073.96 & 11.68 & 0.0086\\
UN & 14 & 4162.26 & 2067.13 & 1.97 & 0.8536\\
AR1 & \phantom{0}6 & 4158.47 & 2073.24 & 10.24 & 0.0166\\
ARH1 & \phantom{0}9 & 4154.23 & 2068.12 & \multicolumn{1}{c}{--} &
\multicolumn{1}{c@{}}{--}\\[3pt]
\multicolumn{6}{c}{Thomas clutter type} \\
CS & \phantom{0}6 & 4197.73 & 2092.86 & 42.57 & <\!0.0001\\
UN & 14 & 4167.06 & 2068.03 & 7.10 & 0.2136\\
AR1 & \phantom{0}6 & 4197.64 & 2092.82 & 42.48 & <\!0.0001\\
ARH1 & \phantom{0}9 & 4161.16 & 2071.58 & \multicolumn{1}{c}{--} &
\multicolumn{1}{c@{}}{--}\\[3pt]
\multicolumn{6}{c}{Hardcore clutter type} \\
CS & \phantom{0}6 & 5244.09 & 2616.04 & 42.63 & <\!0.0001\\
UN & 14 & 5225.46 & 2594.73 & \multicolumn{1}{c}{--} & \multicolumn
{1}{c@{}}{--}\\
AR1 & \phantom{0}6 & 5246.43 & 2617.21 & 44.97 & <\!0.0001\\
ARH1 & \phantom{0}9 & 5228.94 & 2605.47 & 21.47 & 0.0107\\[3pt]
\multicolumn{6}{c}{Strauss clutter type} \\
CS & \phantom{0}6 & 6399.31 & 3193.66 & 66.04 & <\!0.0001\\
UN & 23 & 6353.29 & 3153.65 & 13.97 & 0.4519\\
AR1 & \phantom{0}6 & 6398.92 & 3193.46 & 65.66 & <\!0.0001\\
ARH1 & \phantom{0}9 & 6339.26 & 3160.63 & \multicolumn{1}{c}{--} &
\multicolumn{1}{c@{}}{--}\\
\hline
\end{tabular*}
\end{table}

%
\begin{figure}
\centering
\begin{tabular}{@{}c@{\hspace*{4pt}}c@{}}

\includegraphics{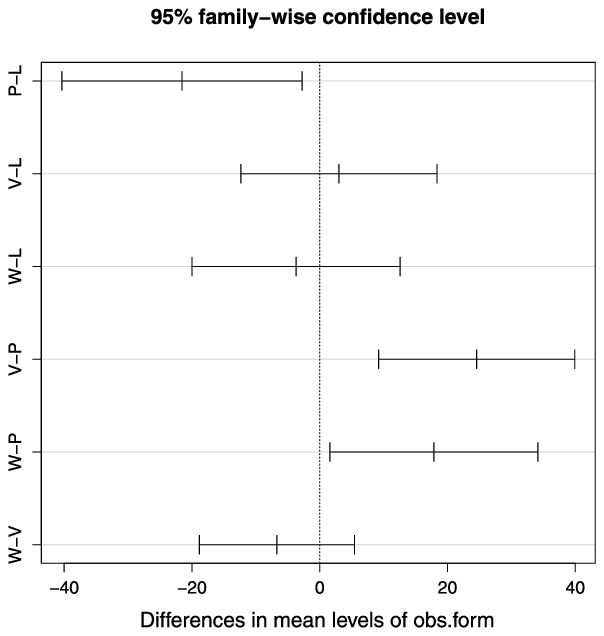}
 & \includegraphics{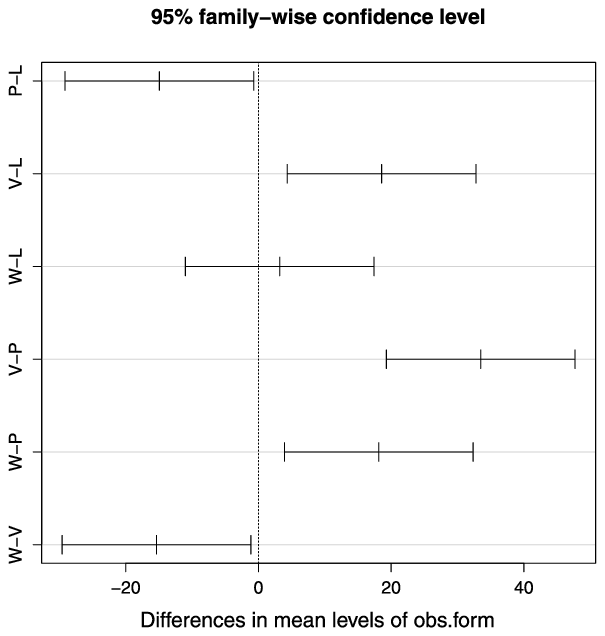}\\
\footnotesize{(a) CSR clutter: P${}={}$154.36,} & \footnotesize{(b)
Inhomogeneous Poisson clutter: P${}={}$155.46,}\\
\footnotesize{L${}={}$175.93, V${}={}$178.93, W${}={}$172.22} &
\footnotesize{L${}={}$170.40, V${}={}$188.96, W${}={}$173.60}\\[5pt]

\includegraphics{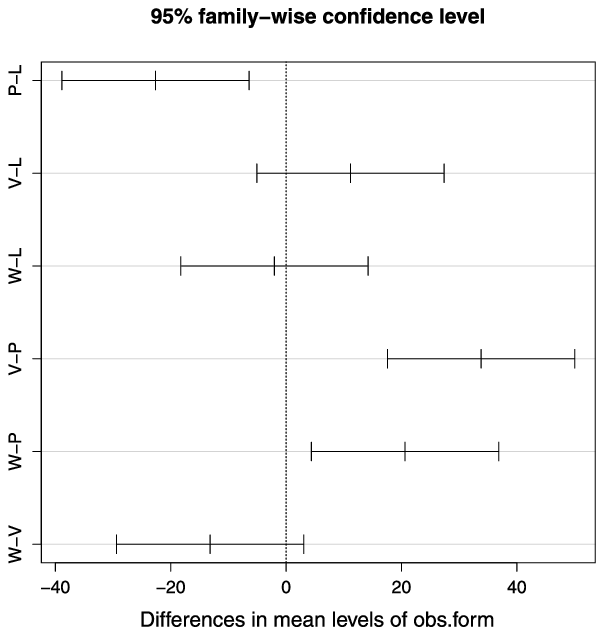}
 & \includegraphics{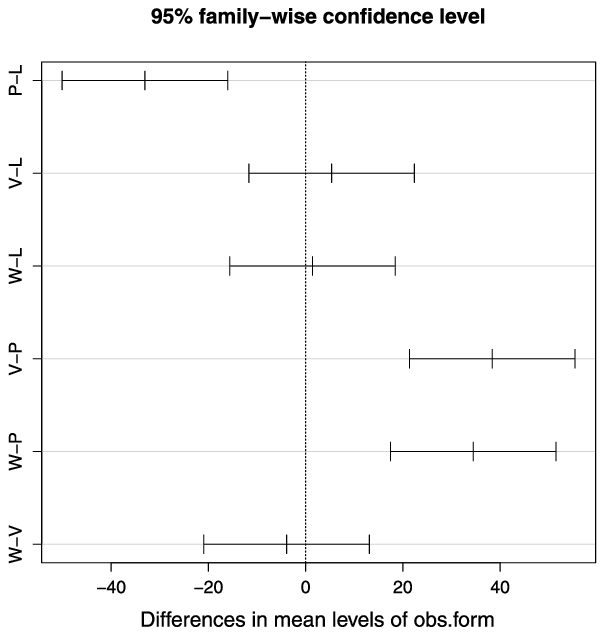}\\
\footnotesize{(c) Mat\'{e}rn clutter: P${}={}$150.59,} & \footnotesize
{(d) Thomas clutter: P${}={}$143.09,}\\
\footnotesize{L${}={}$173.23, V${}={}$184.39, W${}={}$171.21} &
\footnotesize{L${}={}$176.15, V${}={}$181.48, W${}={}$177.56}\\[5pt]
\end{tabular}
\caption{The 95\% family-wise confidence intervals
on the mean differences in traversal lengths based on Tukey's HSD method
for the best performing obstacle type-obstacle number combinations
(written as obs. form)
at each background clutter type.
The average travel lengths for each obstacle form is also provided
below the figures.}\label{figTukeyHSD}
\end{figure}

\setcounter{figure}{5}
\begin{figure}
\centering
\begin{tabular}{@{}c@{\hspace*{4pt}}c@{}}

\includegraphics{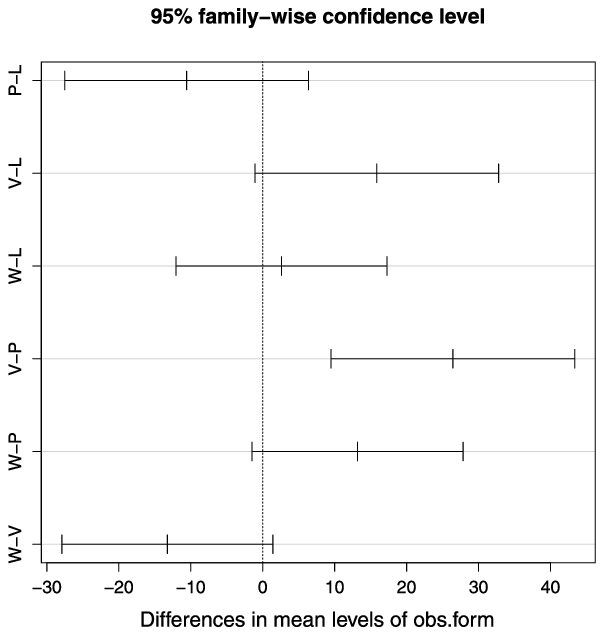}
 & \includegraphics{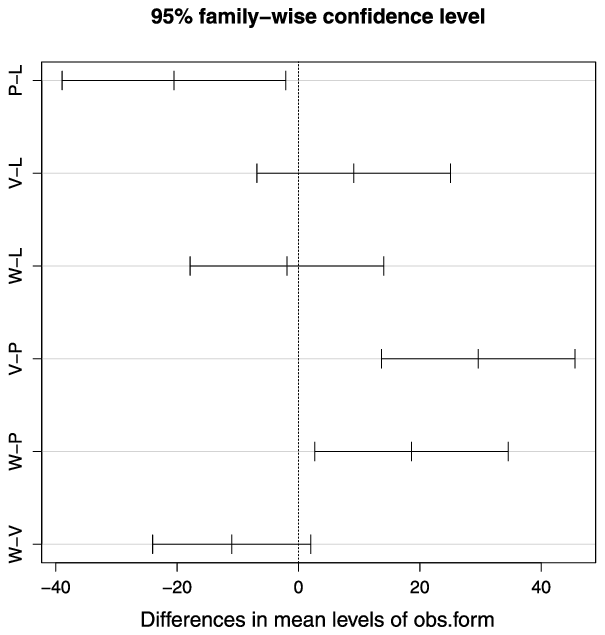}\\
\footnotesize{(e) Hardcore clutter: P${}={}$163.86,} & \footnotesize
{(f) Strauss clutter:
P${}={}$158.94,}\\
\footnotesize{L${}={}$174.43, V${}={}$190.29, W${}={}$177.04} &
\footnotesize{L${}={}$179.45, V${}={}$188.56, W${}={}$177.56}
\end{tabular}
\caption{(Continued).}\vspace*{-2pt} 
\end{figure}

The longest traversal lengths (that are significantly larger than
the others) at each clutter type among the best performers are
marked in bold face in Table~\ref{tabbest-perform}. For each
clutter type, we compare the mean traversal lengths of the best
performer obstacle type-obstacle number combinations by Tukey's
HSD (honestly significant difference) method on mean differences
[\citet{miller1981}]. The corresponding 95\% family-wise confidence
intervals (CI) are plotted in Figure~\ref{figTukeyHSD}, where the
intervals that intersect the vertical line at zero indicate that the
corresponding treatments are not significantly different at the 0.05
level. Best performers for each clutter type are described below.\vspace*{6pt}

 \textit{CSR clutter}:
The longest lengths (that are significantly larger than others)
among best performers (in decreasing order) are at V-shaped, linear
and W-shaped obstacle forms. That is, the lengths for the V-shaped,
linear and W-shaped best performers are not significantly different
from each other at the 0.05 level, although\vadjust{\goodbreak} the mean difference
between V-shaped and W-shaped best performers has $p$-value 0.0645.
Hence, for the CSR clutter, we recommend the use of V80:50,
V50:40, V70:50 or L20:50 obstacle type in this order.
That is, if the cost of the above obstacle placements is about the same,
then any one of them can be used, but V80:50 has a slight advantage,
otherwise the one with the lowest cost is recommended.

 \textit{Inhomogeneous Poisson clutter}:
The longest length among best performers is at V-shaped obstacle
forms. Hence, V90:60 obstacle type is recommended.

 \textit{Mat\'{e}rn clutter}:
The longest length among best performers is at V-shaped obstacle
forms. Hence, V90:50 obstacle type is recommended.

 \textit{Thomas clutter}:
The longest lengths among best performers (in decreasing order) are
at V-shaped, W-shaped and linear obstacle forms. That is, the
lengths for the V-shaped, W-shaped and linear best performers are
not significantly different from each other at the 0.05 level.
Hence, V90:60, W60:60 or L20:50 obstacle types are recommended
in this order.
If there are no cost restrictions,
V90:60 has a slight advantage,
otherwise the one with cheapest construction can be employed.

 \textit{Hardcore clutter}:
The longest length among best performers is at V-shaped obstacle
forms. Hence, V90:50 obstacle type is recommended.

 \textit{Strauss clutter}:
The longest lengths among best performers (in decreasing order) are
at V-shaped and linear obstacle forms, although the mean
difference between V-shaped and linear best performers has
$p$-value 0.0745.
Hence, V80:40 or V90:50 obstacle types are
recommended in this order
as in the previously discussed sense.\vspace*{6pt}

%
\begin{table}
\caption{Cross-tabulation of best performers among obstacle types for
clutter type-obstacle number combinations.
The corresponding mean traversal lengths are provided in
parentheses}\label{tabbest-perf-cross-tab}
\begin{tabular*}{\textwidth}{@{\extracolsep{\fill}}lccccc@{}}
\hline
& \multicolumn{5}{c@{}}{\textbf{Obstacle number}} \\[-4pt]
& \multicolumn{5}{c@{}}{\hrulefill} \\
\textbf{Clutter} & \textbf{20} & \textbf{30} & \textbf{40} & \textbf
{50} & \textbf{60} \\
\hline
CSR & L20 (147.92) & L20 (171.57) & V70 (177.02) & V70 (178.68) & V90 (177.02)\\
& & & V80 (177.84) & V80 (179.12) &
 \\
& & & V50 (178.98) & & \\[3pt]
Inhom. &L30 (151.74) & V60 (167.72) &V60 (175.72) & V90 (184.44) & V90 (188.96)\\
\quad Poisson &  & L30 (168.29) &  & &
 \\
& & L40 (168.85) & & & \\[3pt]
Mat\'{e}rn & L20 (139.07) & L20 (163.10) & L20 (173.23) & V90 (184.39) &
V90 (177.55)\\
& & L30 (164.52) & & & \\[3pt]
Thomas & L20 (141.19) & L20 (166.20) & V70 (178.57) & V50 (179.40) &
V90 (181.48)\\[3pt]
Hardcore & L40 (156.27) & L20 (179.43) & V80 (187.57) & V90 (190.29) &
V60 (180.44)\\[3pt]
Strauss & L60 (143.81) & L20 (177.31) & V80 (188.77) & V90 (188.35) &
V80 (180.37)\\
& L50 (144.97) & & & & \\
\hline
\end{tabular*}
\end{table}

We also provide a cross-tabulation of best performer obstacle type for
each clutter type-obstacle number combination
in Table~\ref{tabbest-perf-cross-tab}.
For example, for the Mat\'{e}rn clutter with 40 obstacles,
the highest traversal length occurs for L20 obstacle pattern.
That is, when we know that the clutter is of Mat\'{e}rn type
and has only 40 obstacles,
the optimal strategy is to employ L20 obstacle scheme.
However, if obstacle number is not restricted (i.e., 60 or more),
the optimal choice is V90 obstacle scheme.

\section{Example data: Maritime minefield application}
\label{seccobra} 

\subsection{Data description}
\label{secdata-desc} 
\begin{figure}
\centering
\begin{tabular}{@{}c@{}}

\includegraphics{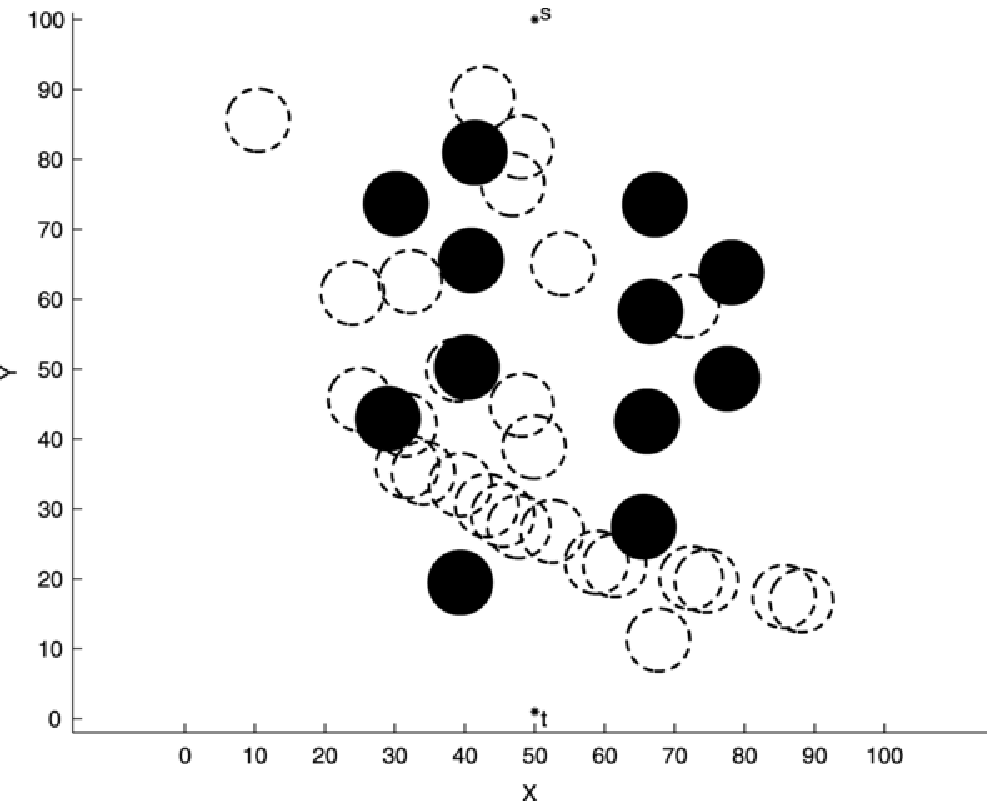}
 \\
\footnotesize{(a) Actual status of the minefield}\\[6pt]

\includegraphics{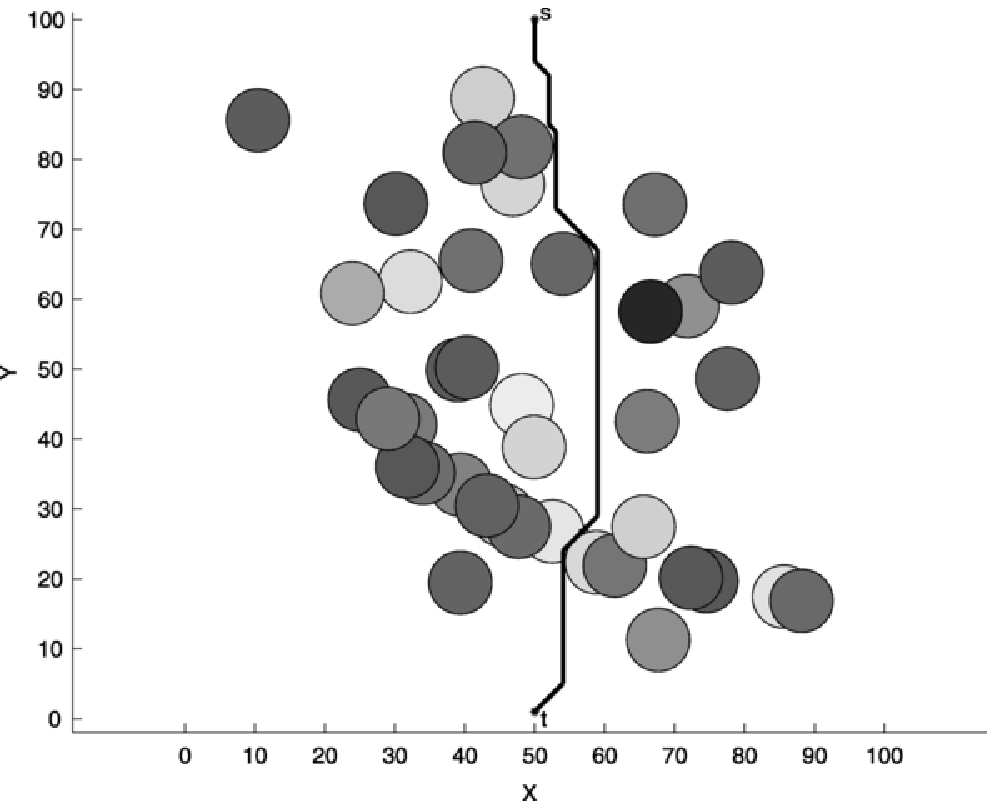}
\\
\footnotesize{(b) The minefield as seen by the NAVA and its traversal}
\end{tabular}
\caption{The COBRA data and the NAVA's traversal using the ARD algorithm.
Figure \textup{(a)} shows the actual allocation of the minefield where
clutter disks are denoted by dashed circles and mines are denoted by
black disks.
Figure \textup{(b)} depicts the COBRA minefield as seen by the NAVA and
its $s-t$ traversal.}\label{figcobra}
\end{figure}

Maritime minefield detection, localization and navigation have
received considerable attention from scientific and engineering
communities recently; see, for example, \citet{wit95}, \citet{ran08}
and the references cited therein. Operational concepts for maritime
minefield detection via unmanned aerial vehicles are discussed in
\citet{wit95} wherein multi-spectral imagery of a potential
minefield is examined and locations of potential mines are
identified using a classification algorithm. Of particular interest is a
U.S. Navy minefield data set (called the COBRA data) that first appeared
in \citet{wit95} and was later referred to in \citet{pri97},
\citet{rdp1}, \citet{rdp2},
\citet{ye10}, \citet{ye11}, and \citet{rdp3}. The COBRA data,
illustrated in Figure~\ref{figcobra}(a), has a total of 39
disk-shaped potential mines of which 27 are clutter and the remaining
12 are
true mines [\citet{ye11}].
The original data coordinates were scaled and shifted so that
clutter disk centers are inside the region $[10,90] \times[10,90]$.
As in our simulations, we take $s=(50,100)$ and $t=(50,1)$,
disk radius as $r=4.5$, and cost of disambiguation as $c=5$ (our
simulation environment was in fact inspired by the COBRA data).
When the ARD algorithm is applied on the COBRA data [shown in Figure
\ref{figcobra}(b)], the actual traversal length is 111.46
units with one disambiguation.

%

%
\begin{figure}
\centering
\begin{tabular}{@{}cc@{}}

\includegraphics{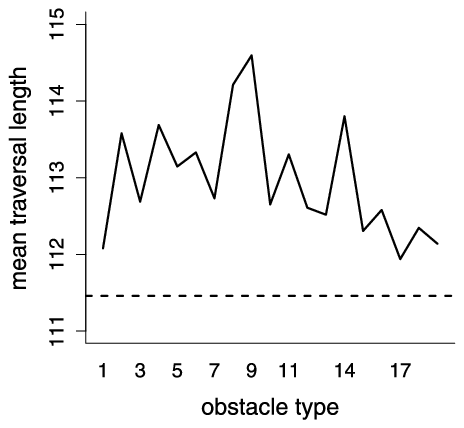}
 & \includegraphics{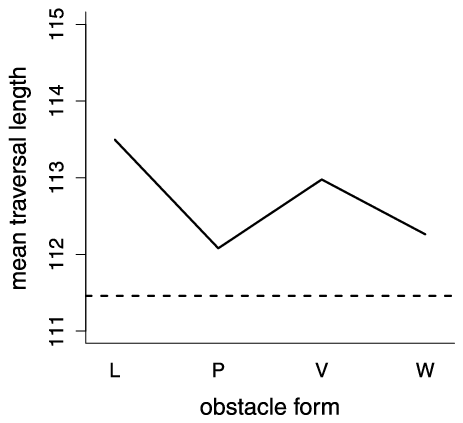}
\end{tabular}
\caption{Plots for mean traversal lengths for the 19 obstacle types (left)
and 4 obstacle forms (right) with 12 mines and the 27 COBRA clutter disks.
The horizontal dashed line is at 111.46 which is the traversal length
for the 12 mines as originally placed in the COBRA data.}\label{figex-12-mines}
\end{figure}

A visual inspection of the COBRA clutter suggests that it does not seem
to fit any one of the six clutter distribution types considered in this
work. Instead, in the scaled coordinates, the pattern looks like a realization
from an inhomogeneous Poisson process with intensity being inversely
proportional to the distance to the diagonal line, $y=-x+100$.
That is, the clutter is more concentrated around this line, compared to
regions further away.

\subsection{Analysis of traversal lengths for the example data with 12 mines}
\label{secex-obs-12mines}

In this section we investigate whether placing the 12 mines in our
example data set using any one of our 19 obstacle placing schemes
results in longer traversal lengths compared to their actual
placements.
Using the 19 obstacle schemes from 4 obstacle forms we consider, mean
traversal lengths are plotted in Figure~\ref{figex-12-mines} with
100 realizations from each scheme. The shortest traversal lengths
occur at W and P obstacle forms with traversal length 112.1, and the highest
traversal lengths occur at linear obstacle type L20 with traversal
length 114.60.
While the difference is not drastic, it is a better strategy on the
average to use the L20 obstacle placing scheme to place these 12
mines compared to their original allocation in the COBRA data.
Since we are using the same (COBRA) clutter realization for each of 1900
Monte Carlo replications,
the setting does not lend itself for repeated measures ANOVA.
There seems to be a dependence between traversal length measurements,
but since the same clutter is used for each realization,
it is as if the same subject receives all 1900 realizations of treatments.
Hence, we use the usual ANOVA in our subsequent analysis.
The ANOVA assumptions are the same as the repeated measures ANOVA
with compound symmetry having zero covariance.
We also find significant differences in mean traversal lengths among
the obstacle types ($p<0.0001$) and the obstacle forms ($p<0.0001$).
Mean traversal lengths are significantly different for the 5\vadjust{\goodbreak}
obstacle types among the W-shaped obstacle form ($p=0.0303$).
However, mean traversal lengths are not significantly different for
the 8 obstacle types among the linear obstacle form
($p=0.1287$) nor the 5 obstacle types among the V-shaped obstacle
forms ($p=0.1931$).

%
\begin{figure}

\includegraphics{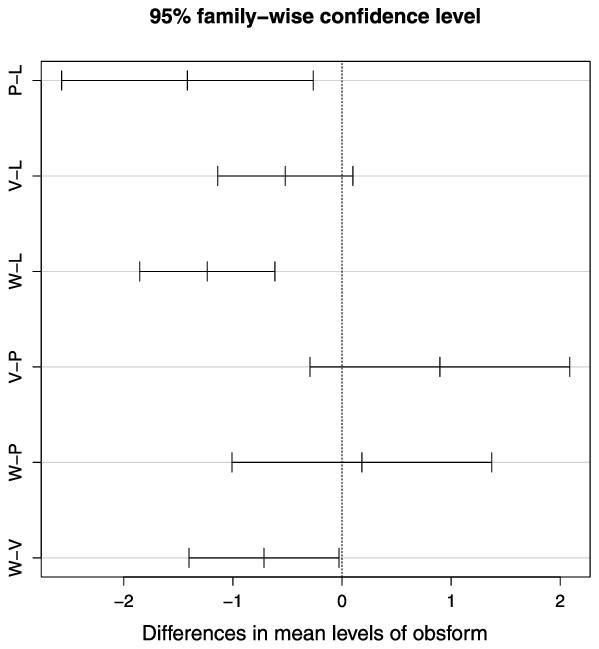}

\caption{The 95\% family-wise confidence intervals
on the mean differences in traversal lengths based on Tukey's HSD method
for the obstacle types with the COBRA clutter.
The mean traversal lengths for the treatment combinations are
P${}={}$112.08, W${}={}$112.26, V${}={}$112.96 and
L${}={}$113.50.}\label{figex-TukeyHSD-12mines}
\end{figure}

We now compare the mean traversal lengths for the obstacle forms
by Tukey's HSD method on mean differences.
The corresponding 95\% family-wise confidence
intervals are plotted in Figure~\ref{figex-TukeyHSD-12mines}.
We observe that mean traversal lengths are not significantly different
for linear and V-shaped obstacle forms ($p=0.1363$),
but linear form is significantly better than the P obstacle form
($p=0.0087$) and W-shaped obstacle form ($p<0.0001$). The
P obstacle form is not significantly different from the V-shaped form
($p=0.2127$) nor from the W-shaped form ($p=0.9793$).
On the other hand, the V-shaped obstacle form is significantly better
than the W-shaped form ($p=0.0378$).

We also experiment with our 19 obstacle
placement schemes with a different number of mines (ranging from 20 to
60 in 10 unit increments) to gain insight
into which scheme(s) perform better for the example (COBRA) data
clutter disks.
Again,
since we are using the same COBRA clutter as the background pattern
for each Monte Carlo realization of obstacle type-obstacle number combination,
we use the usual ANOVA in our analysis.

\subsection{Overall comparison of traversal lengths for the example data}
\label{secex-overall-length}
\begin{figure}

\includegraphics{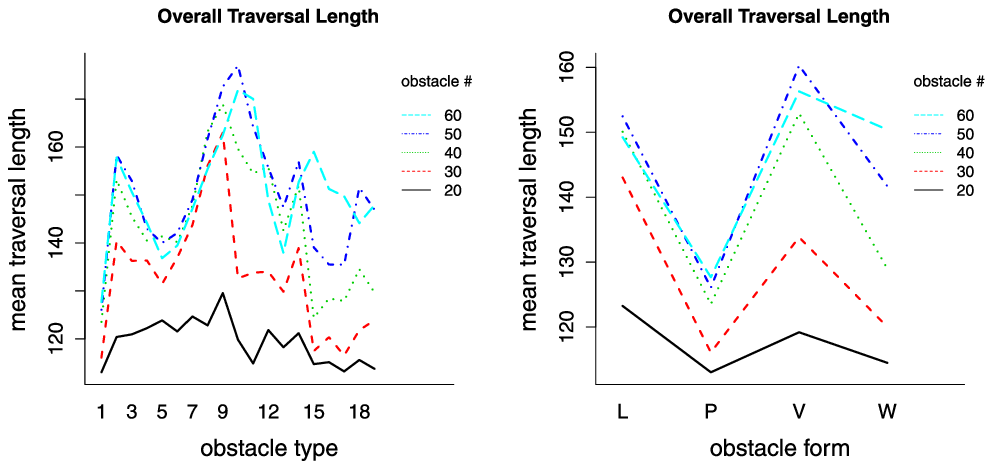}

\caption{The profile plots for obstacle type/form versus obstacle number
for the example clutter
pattern.}\label{figex-overall-interaction}\vspace*{6pt}
\end{figure}

We first consider the following model:
%
\begin{equation}
\label{eqnex-overall}
T_{ijk}=\mu_0+\mu^O_i+\mu^{\mathrm{NO}}_j+\mu^{O,\mathrm{NO}}_{i,j}+\varepsilon_{ijk},
\end{equation}
where
$\mu_0$ is the overall mean,
$\mu^O_i$ is the mean for obstacle type $i$,
$\mu^{\mathrm{NO}}_j$ is the mean for obstacle number level $j$,
$\mu^{O,\mathrm{NO}}_{i,j}$ is the mean for the obstacle type $i$ and obstacle
number level $j$ combination
(which stands for the interaction between these factors),
$T_{ijk}$ is the $k${th} traversal length
for obstacle type $i$ and obstacle number level $j$,
and
$\varepsilon_{ijk}$ is the error term
with
$i=1,2,\ldots,19$,
$j=1,2,\ldots,5$, and
$k=1,2,\ldots,100$.
We find significant interaction between obstacle type and obstacle
number levels ($p<0.0001$), and between obstacle form and obstacle
number levels ($p<0.0001$), which means the trends in mean lengths
plotted in Figure~\ref{figex-overall-interaction} are
significantly nonparallel. Hence, it is not reasonable to compare
the mean traversal lengths for obstacle types/forms and obstacle
number levels (i.e., for the main effects of obstacle type/forms and
obstacle numbers), but instead, for example, it will make sense to
compare the mean length values for obstacle number levels at each
obstacle type or form.

%

%
\begin{table}
\tabcolsep=0pt
\caption{The shortest and
longest traversal lengths and the corresponding\break treatment types for
overall comparisons, and comparisons\break at specific clutter types,
obstacle types and obstacle numbers
}\label{tabex-shortest-highest-lengths}
\begin{tabular*}{\textwidth}{@{\extracolsep{\fill}}lcccc@{}}
\hline
& \multicolumn{2}{c}{\textbf{Shortest}} &
\multicolumn{2}{c@{}}{\textbf{Longest}}\\[-4pt]
& \multicolumn{2}{c}{\hrulefill} & \multicolumn{2}{c@{}}{\hrulefill}\\
& \textbf{Traversal} & \textbf{Treatment} & \textbf{Traversal} & \textbf
{Treatment}\\
& \textbf{length (s)} & \textbf{type (s)} & \textbf{length (s)} &
\textbf{type (s)}\\
\hline
\multicolumn{5}{c}{Overall} \\
& 113.03, 113.21, & P:20, W70:20, & 176.99 & V90:50\\
& 113.76 & W50:20 & & \\[6pt]
\multicolumn{5}{c}{Obstacle form} \\
P & 113.03 & 20 & 126.07, 127.65 & 50, 60 \\
Linear & 120.40, 120.94 & L90:20, L80:20 & 172.56 & L20:50\\
V-Shaped & 114.88 & V80:20 & 176.99 & V90:50\\
W-Shaped & 113.21, 113.76 & W70:20, W50:20 & 159.00 & W90:60 \\[6pt]
\multicolumn{5}{c}{Number of obstacles} \\
20 & 113.03, 113.21, & P, W70, & 129.54 & L20 \\
& 113.76 & W50 & & \\
30 & 116.10, 116.52 & P, W70, & 162.97 & L20\\
& 117.47 & W90 & & \\
40 & 123.54, 124.38 & P, W90 & 168.91 & L20\\
50 & 126.07 & P & 176.99 & V90 \\
60 & 127.67 & P & 170.02, 171.92 & V80, V90 \\
\hline
\end{tabular*}
\end{table}

The trend in traversal lengths as number of mines increases is similar
to our simulation results. At P and W-shaped obstacle forms, traversal
lengths tend to increase as the obstacle number increases. On the other
hand, at linear and V-shaped obstacle forms, traversal lengths exhibit
a concave-down trend. For linear windows, the shortest length occurs at
20 mines and the longest occurs at 50 mines. For V-shaped windows,
shortest length occurs at 20 mines, and longest length occurs at 50
mines. For 20--30 mines, the longest traversal lengths occur at linear
obstacle forms, and for 40--60 mines,
the longest lengths occur at V-shaped obstacle forms.

The shortest and longest traversal length performances are presented
in Table~\ref{tabex-shortest-highest-lengths}. Overall, the shortest
length is about 113 units which occurs at P:20, W70:20 and W50:20
treatment combinations, and the longest length is about 177 which
occurs at V90:50 treatment combination.

\subsection{Analysis of traversal lengths at each obstacle form for the
example data}
\label{secobs-forms}
\begin{figure}
\centering
\begin{tabular}{@{}cc@{}}

\includegraphics{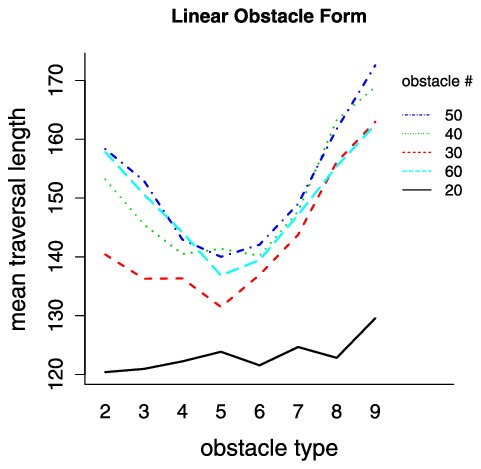}
 & \includegraphics{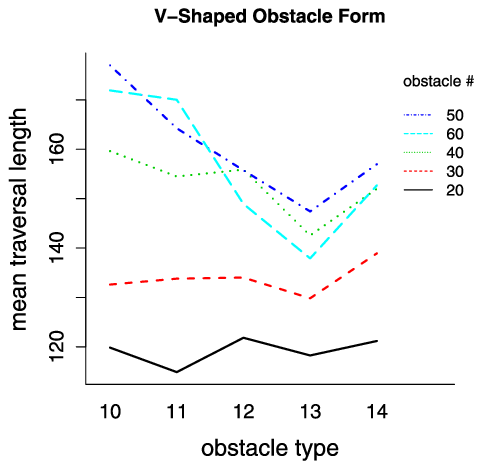}\\
\footnotesize{(a)} & \footnotesize{(b)}\vspace*{6pt}
\end{tabular}
\centering
\begin{tabular}{@{}c@{}}

\includegraphics{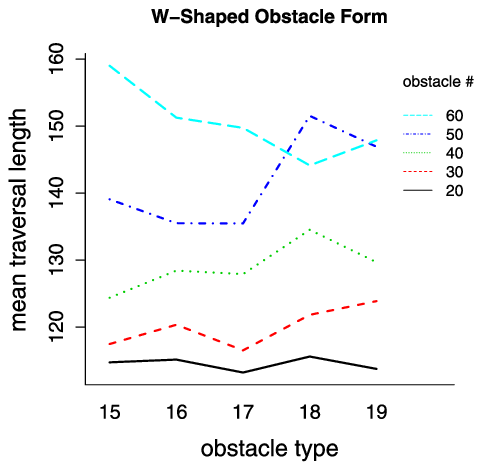}
\\
\footnotesize{(c)}
\end{tabular}
\caption{The profile plots for obstacle type versus obstacle number level
at obstacle forms other than P.}\label{figex-OBS-forms-ObsvsMine-interaction}
\end{figure}

We now investigate the interaction between obstacle type and obstacle
number at each obstacle form. Note that no interaction is well defined
for the P obstacle form. For other obstacle forms such interaction is
possible. In Figure~\ref{figex-OBS-forms-ObsvsMine-interaction} we
present the profile plots for interaction between obstacle type and
obstacle number at each obstacle form (other than the P obstacle form).

At the P obstacle form,
we consider the following model:
%
\begin{equation}
\label{eqnex-CSR-Obs-model}
T_{1jk}=\mu_0+\mu^O_1+\mu^{\mathrm{NO}}_j+\varepsilon_{1jk}.
\end{equation}
At other obstacle forms,
we consider the following model:
%
\begin{equation}
\label{eqnex-Obs-Form-nonCSR-model}
T_{ijk}=\mu_0+\mu^O_i+\mu^{\mathrm{NO}}_j+\mu^{O,\mathrm{NO}}_{i,j}+\varepsilon_{ijk},
\end{equation}
where obstacle form indices $j=2,3,4$
stand for linear, V-shaped and W-shaped obstacle forms, respectively.\vspace*{6pt}

%

 \textit{The P obstacle form}:
The traversal lengths are significantly different between obstacle
number levels
($p<0.0001$).

 \textit{Linear obstacle form}:
We find significant interaction between obstacle type and obstacle
number levels ($p<0.0001$). Hence, the trends in mean lengths plotted in
Figure~\ref{figex-OBS-forms-ObsvsMine-interaction}(a) are significantly
different from being parallel. So, it is not reasonable to test for
the main effects of obstacle types or obstacle
number levels at the linear obstacle form.

 \textit{V-shaped obstacle form}:
We find significant interaction between obstacle type and obstacle
number levels ($p<0.0001$). Hence, the corresponding trends in mean
length plotted in Figure~\ref{figex-OBS-forms-ObsvsMine-interaction}(b)
are significantly nonparallel.
So, it is not reasonable to compare the main effects
of obstacle types and obstacle number levels.

 \textit{W-shaped obstacle form}:
We find significant interaction between obstacle type and obstacle
number levels ($p=0.0004$).
Hence, the corresponding trends in mean
length plotted in Figure~\ref{figex-OBS-forms-ObsvsMine-interaction}(c)
are significantly nonparallel. So, it is not reasonable to compare
the main effects of obstacle type and obstacle number levels here.

Analysis of traversal lengths for each obstacle form is given below. The
shortest and longest traversal lengths together with the
corresponding treatment combinations are presented in Table
\ref{tabex-shortest-highest-lengths}.\vspace*{6pt}

 \textit{The P obstacle form}:
The shortest length is about 113 which occurs at 20 mines,
and the longest length is about 127
which occurs at (50 or 60) mines.
On the average, traversal lengths tend to increase as the obstacle
number increases.

 \textit{Linear obstacle form}:
The shortest length is about 120 which occurs at L90:20 and L80:20
treatment types, and the longest length is about 173 which
occurs at L20:50 treatment type.
At each obstacle type,
the shortest lengths occur at 20 obstacles
and the highest lengths tend to occur at 50 obstacles.
For 20 mines,
as the obstacle type level increases from 2 to 9
(i.e., as distance to coast decreases),
mean traversal length tends to increase slightly.
However, for other mine number levels,
the trend exhibits a concave-up behavior
(i.e., first decreases, reaches a minimum and then increases).

 \textit{V-shaped obstacle form}:
The shortest length is about 115 which occurs at V80:20
treatment type, and the longest length is about 177 which
occurs at V90:50 treatment type.
As the V-shaped pattern changes from 10--14 (i.e., as distance to coast
decreases),
the traversal lengths tend to stay stable for 20--30 mines,
but for 40--60 mines, it follows a concave-up behavior.

 \textit{W-shaped obstacle form}:
The shortest length is about 113 which occurs at W70:20, W50:20
treatment types, and the longest length is about 159 which
occurs at W90:60 treatment type.
Traversal lengths tend to stay stable for 20--40 mines,
but for 50 mines it exhibits a concave-down behavior,
and for 60 mines a concave-up behavior.

\subsection{Analysis of traversal lengths at each obstacle number level
for the example data}
\label{secex-obstacle-number}
We also investigate the relation between traversal lengths and
obstacle type/form at each obstacle number level.
At each obstacle number level,
we consider a model as in equation \eqref{eqnex-overall}.
For example, with 20 obstacles,
the model is
%
\begin{equation}
\label{eqn20Mines-model}
T_{i1k}=\mu_0+\mu^O_i+\mu^{\mathrm{NO}}_1+\varepsilon_{i1k}.
\end{equation}
Comparison of traversal lengths for each number of
obstacles is presented below. The shortest and longest traversal
lengths together with the corresponding treatment combinations are
presented in Table~\ref{tabshortest-highest-lengths}.

For 20 mines, the shortest traversal length is about 113 which occurs
at~P, W70, W50
treatment types, and the longest length is about 129 which
occurs at L20 treatment type.
For 30 mines,
the shortest length is about 116 which occurs at P, W70, W90
treatment types, and the longest length is about 163 which
occurs at L20 treatment type.
For 40 mines,
the shortest length is about 124 which occurs at P, W90 treatment
types, and the longest length is about 168 which occurs at L20
treatment type.
For 50 mines,
the shortest length is about 126 which occurs at P treatment
type, and the longest length is about 177 which occurs at V90
treatment type.
For 60 mines,
the shortest length is about 128 which occurs at P treatment
type, and the longest length is about 171 which occurs at V80, V90
treatment types.

For 20 and 30 mines,
on the average,
the longest travel lengths occur for the linear obstacle form.
On the other hand,
for 40--60 mines,
the longest traversal lengths occur for the V-shaped obstacle form.
For 20--60 mines,
the shortest lengths occur for the P obstacle form.
Furthermore, for 20--60 mines,
the traversal lengths have a concave-up trend as distance to coast decreases.

\subsection{Comparison of best performers for the example data}
\label{seccomp-best-perf}
The overall best performer and best performers for each
obstacle type are presented in Table
\ref{tabex-shortest-highest-lengths}.
Since there are multiple
best performer obstacle type-obstacle number combinations,
we compare the traversal lengths for obstacle form levels.
We consider the following model:
%
\begin{equation}
\label{eqnex-best-perf-model}
T_{ij}=\mu_0+\mu^{\mathrm{OF}}_i+\varepsilon_{ij},
\end{equation}
where
$\mu_0$ is the overall mean,
$\mu^{\mathrm{OF}}_i$ is the main effect of obstacle form $i$,
and
$\varepsilon_{ij}$ is the error term
for $i=1,2,3,4$ (which correspond to P, linear, V-shaped and W-shaped
obstacle forms)
and $j=1,2,\ldots,n_i$,
where $n_i$ is $k\times100$ with $k$ being the number of treatment
combinations that are best performers.
For example, $k=2$ for the P obstacle type.

\begin{figure}

\includegraphics{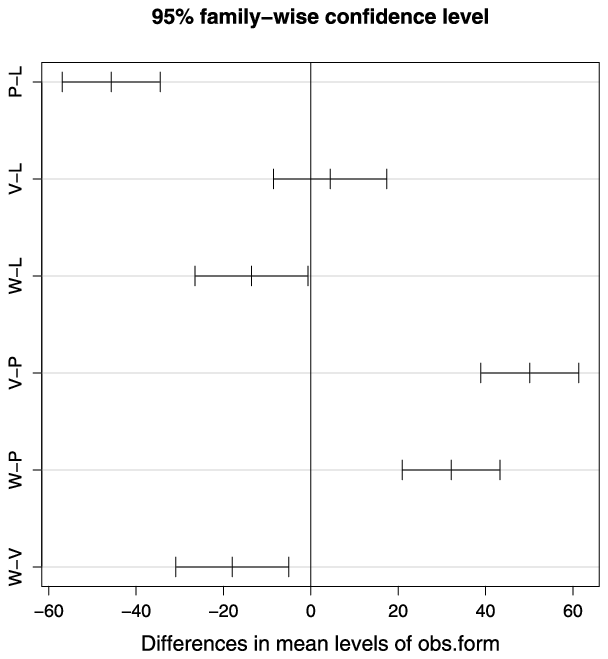}

\caption{The 95\% family-wise confidence intervals
on the mean differences in traversal lengths based on Tukey's HSD method
for the best performing obstacle type-obstacle number combinations
at each background clutter type.
The mean traversal lengths for the treatment combinations are
P${}={}$126.86, L${}={}$172.56, V${}={}$176.99 and W${}={}$159.00.}\label{figex-TukeyHSD}
\end{figure}

We compare the mean traversal lengths of the best
performer obstacle type-obstacle number combinations by Tukey's
HSD method on mean differences.
The corresponding 95\% family-wise confidence
intervals are plotted in Figure~\ref{figex-TukeyHSD}.
Notice that best performers for linear and V-shaped obstacle forms are
not significantly different ($p=0.8148$),
but the V-shaped is significantly larger than P ($p<0.0001$) and
W-shaped obstacle forms ($p=0.0021$).
Furthermore, the
linear is also significantly larger than P ($p<0.0001$) and W-shaped
obstacle forms ($p=0.0358$).
Finally, the W-shaped form is significantly larger than the P obstacle
form ($p<0.0001$).

\section{Discussion and conclusions}
\label{secdisc} 

In this work we introduce the obstacle placement with the
disambiguations (OPD) problem wherein the objective is to place a
given number of true obstacles in between the clutter so as to
maximize the traversal length of the navigating agent (NAVA) in a
game-theoretic sense. We consider a specific version of the problem
where the obstacle placing agent (OPA) knows the clutter type (i.e., the
clutter spatial distribution), but not the exact location of clutter
disks. We investigate relative efficiency of a variety of obstacle
placement patterns against different background clutter types.
Our goal is to explore the effect of the number of obstacles on the NAVA's
traversal length and to determine which obstacle placement patterns
perform better for a given clutter type.
We also present an extensive case study on a real-world maritime
minefield data set. We believe that such an analysis within a maritime
minefield context has a significant potential in the design of more
efficient and cost-effective
interdiction systems.

Our setup leads to a three-way repeated measures ANOVA
problem where the treatment factors are the clutter type,
number of obstacles and the obstacle placement pattern, with the
response variable being the NAVA's total traversal length.
We choose repeated measures ANOVA instead of the usual ANOVA to
gain more precision and power in our analysis.
Furthermore,
the model assumptions for repeated measures ANOVA are satisfied
with the flexibility of modeling different types of correlation
between repeated measures.
We consider a
total of 6 clutter types: homogeneous Poisson process (also
known as CSR), inhomogeneous Poisson
process, Mat\'{e}rn, Thomas, hard-core and Strauss point
processes. We consider 5 different numbers of obstacles (20, 30$,\ldots, $60), and we experiment with a total of 19 different
obstacle placement patterns sampled from CSR in four different
forms: the clutter sampling window $P$, linear, V- and W-shaped
polygon windows.

Extensive statistical analysis of our Monte Carlo simulations
indicate that as the clutter spatial distribution becomes more regular
(clustered),
the traversal length gets longer (shorter). In terms of obstacle number levels,
the traversal length tends to follow a concave-down trend (i.e.,
increases, reaches a peak and then decreases) as the number of
obstacles increases. The reason for the traversal trend following
such a concave-down trend is that the traversal length tends to
increase to a certain extent as the hindrance disk
(obstacle${}+{}$clutter) density increases, reaches an optimum and then
decreases, as the NAVA tends to avoid the obstacle window altogether
when the hindrance density becomes too high. This is perhaps a
counterintuitive result at first, as placing more and more obstacles
in the obstacle field becomes detrimental after a certain point from the
OPA's perspective.

In terms of obstacle forms, the shortest traversals tend to occur
for the P obstacle form. The longest traversal lengths, on the other
hand, tend to occur for the V-shaped obstacle form. It appears that
the V-shaped obstacle form tends to trap the NAVA within its elbow
like (convex) region, especially in the presence of a large number
of obstacles.
Under CSR clutter,
linear obstacle forms enjoy the longest traversal for small obstacle
numbers (20--30),
and V-shaped obstacle forms enjoy the longest traversal for moderate to
large obstacle numbers (40--60).
For other clutter types, the
V-shaped obstacle form that is closer to the starting point with a
large number of obstacles tends to result in the longest traversals.
In particular,
the longest traversals occur at V90:60 for the CSR and Thomas clutter types;
V90:50 for Mat\'{e}rn, hardcore, and Strauss clutter types;
and V70:50 for the inhomogeneous Poisson clutter type.
Among linear obstacle forms,
the traversal tends to get longer as the linear obstacle window gets
closer to the target,
while among V- and W-shaped obstacle forms traversal tends to get longer
as the obstacle window gets further away from the target.
Thus, with a small number of obstacles (i.e., 20--30 obstacles),
the best performers are linear obstacle windows closer to the target.
As for larger obstacle numbers (i.e., 40--60 obstacles),
the best performers are V-shaped obstacle windows closer to the
starting point.

Our results and conclusions are valid only for the
specific experimental setup we consider,
so they are likely to
change for different clutter and/or obstacle windows, mark
distribution, disambiguation cost or clutter type parameters.
Nonetheless, the statistical analysis we present in this
study can easily be adapted to analyze OPD problem instances within
such different environments.
In fact,
we also study a real-world maritime minefield data set with 27 clutter
and 12 actual obstacles
(see Section~\ref{seccobra}).
Even though the real-world clutter pattern does not resemble
any of the clutter patterns we consider in our simulations,
analysis of our obstacle placement schemes for the
real-world clutter results in a similar conclusion as our Monte Carlo
simulations:
for smaller obstacle numbers,
larger traversal lengths occur for linear obstacle forms closer to the target,
and
for moderate-to-larger obstacle numbers,
larger traversal lengths occur for V-shaped obstacle forms closer to
the starting point.
In what follows, we provide a brief discussion on several issues
related to our research.

 \textit{Asymmetry of information}: An inherent assumption in our framework
is asymmetry of information between the NAVA and the OPA: it is assumed
that the
OPA knows the distribution of the clutter whereas the NAVA does not. On
the other hand, should the NAVA have certain prior information on
clutter distribution, it can incorporate this information into its
traversal strategy by updating disk marks accordingly. Specifically, the
NAVA can assign lower marks to disks fitting the overall clutter
pattern while assigning higher marks to disks that do not
(perhaps when considered in conjunction with spectral image
properties of the disks). Thus, our framework allows for
incorporation of any information on clutter distribution from the NAVA's
perspective. In fact, our simulation setup can be seen as accounting
for this asymmetry of information to a certain extent: clutter marks
are sampled from $\operatorname{Beta}(6,2)$ (with a mean of 0.25) whereas obstacle
marks are sampled from $\operatorname{Beta}(2,6)$ (with a mean of 0.75).

 \textit{Sampling obstacle centers from homogeneous Poisson distribution}:
In our simulations, obstacle disk centers are sampled from a
homogeneous Poisson distribution within their respective obstacle
windows, potentially resulting in overlapping obstacles. On the
other hand, it can be argued that this is not an ideal strategy, as
it is more sensible for the OPA to maximize the space occupied by true
obstacles. From that perspective, a hardcore process should have
been preferred for sampling obstacle disk centers. The reason we
chose homogeneous Poisson over hardcore was to keep information
asymmetry at a minimum. Specifically, in the case of hardcore obstacle
centers, if a disambiguated disk turns out to be a true obstacle,
the optimal strategy for the NAVA would be to decrease marks of surrounding
disks in some fashion. On the other hand,
with a homogeneous
Poisson, as in our simulations, learning that a disambiguated disk
is a true obstacle does not give the NAVA any additional information
regarding the actual status of surrounding disks. The crucial
observation here is that the ARD algorithm, as currently
implemented, would not have accounted for such mark dependencies had
we used hardcore instead of homogeneous Poisson, giving the NAVA an
unfair disadvantage. We leave it to future research to adapt the ARD
algorithm for such a dependency structure and then use the hardcore
pattern for generating obstacle disk centers.

 \textit{Limitations of the ARD algorithm}:
The ARD algorithm is currently the state-of-the-art method for
optimal navigation in stochastic environments in the presence of a
disambiguation capability. On the other hand, it is merely a
heuristic method with no guaranteed performance bounds---yet the
underlying problem is a challenging stochastic optimization problem
and all known exact methods have exponential computational
complexity. Our observation that average traversal length tends to
be concave-down with respect to the number of obstacles might indeed
be attributed to short-comings of the ARD algorithm rather than
benefits of fewer obstacles. This issue warrants further
investigation of the ARD algorithm's performance, which is left to
future research.\vspace*{6pt}

Our work can be extended in several directions. First, more clutter
types and various other obstacle placement schemes can be
considered. Second, another variant of the OPD problem can be
studied where the OPA knows the exact locations of background
clutter disks prior to placing the true obstacles.
In this particular case, the OPA can strategically place the obstacles (as
opposed to placing them randomly inside a predetermined obstacle
window). This approach is likely to be more efficient in terms of
slowing down the progress of the NAVA. However, as mentioned earlier,
one downside of this specific variant is that it requires that the OPA
knows the sensor technology of the NAVA, that is, the OPA has
information on
which specific areas NAVA's sensors detect as potential obstacle regions.
Third, a more general version of the OPD problem can be investigated
where the OPA also has control over the clutter disk locations (in
addition to the obstacle disk locations). In this version, the OPA's
challenge would be to place a given number of true \textit{and}
false obstacle disks in the field, again so as to maximize the
traversal length of the NAVA in a game-theoretic sense.

\section*{Acknowledgments}
The authors thank the area editor, the associate editor and two
anonymous referees
whose valuable comments and suggestions greatly improved the
presentation and flow of
this article. The MATLAB code of the ARD algorithm on 8-regular
lattices can be provided upon request.

%


\printaddresses

\end{document}